\documentclass[pre,preprint,superscriptaddress,amsmath,amssymb,floatfix]{revtex4}
\usepackage{graphics}
\usepackage[dvips]{graphicx}
\usepackage{dcolumn} 
\usepackage{bm} 
\begin{document}
\title{Universality Class of the Critical Point in the 
Restricted Primitive Model of Ionic Systems}
 \author { A.Ciach}
\affiliation{Institute of Physical Chemistry,
 Polish Academy of Sciences, 01-224 Warszawa, Poland}
\date{\today}
\begin{abstract}
A coarse-grained description of the restricted primitive model is
considered in terms of the local charge- and number-density
fields. Exact reduction to a one-field theory is derived, and exact
expressions for the number-density correlation functions in terms of
higher-order correlation functions for the charge-density are
given. It is shown that in continuum space the singularity of the
charge-density correlation function associated with short-wavelength
charge-ordering disappears when charge-density fluctuations are
included by following the Brazovskii approach. The related singularity
of the individual Feynman diagrams contributing to the number-density
correlation functions is cured when all the diagrams are segregated
ito disjoint sets according to their topological structure. By
performing a resummation of all diagrams belonging to each set a
regular expression represented by a secondary diagram is obtained. The
secondary diagrams are again segregated into disjoint sets, and the
series of all the secondary diagrams belonging to a given set is
represented by a hyperdiagram. A one-to-one correspondence between the
hyperdiagrams contributing to the number-density vertex functions, and
diagrams contributing to the order-parameter vertex functions in a
certain model system belonging to the Ising universality class is
demonstrated. Corrections to scaling associated with irrelevant
operators that are present in the model-system Hamiltonian, and other
corrections specific to the RPM are also discussed.
\end{abstract}
\maketitle 

\section{Introduction}

The question of universality class of the critical point associated
with a separation into ion-dilute and ion-dense phases in ionic systems
has been a subject of an intensive debate for many years. The studies
focused mainly on the restricted primitive model (RPM) and we shall
focus on the RPM in this work. In the RPM equaly-sized hard spheres
with positive and negative charges of the same magnitute are dissolved
in a structureless solvent (or in vacuum), and the system is globally
charge-neutral. It was predicted theoretically \cite{stell:76:0} that
the RPM can phase separate, and the separation was observed
experimentally and in simulations. Stell put forward arguments that
the RPM critical point belongs to the Ising universality class already
many years ago~\cite{hafskjold:82:0,stell:92:0,stell:95:0}, and the
same conclusion followed later from the mesoscopic description
described in Refs.\cite{ciach:00:0,ciach:02:0,ciach:05:0}.  Earlier
simulation results for the RPM in continuum space were not
sufficiently accurate to allow for definite conclusions concerning the
universality class, however. Moreover, some early experimental results
confirmed the Ising criticality~\cite{japas:90:0,narayanan:94:0},
whereas other results were in conformity with the mean-field
criticality
\cite{pitzer:90:0,singh:90:0,zhang:92:0} or indicated a crossover to the
Ising-type critical exponents unusually close to the critical
point~\cite{weingaertner:01:0,bianchi:01:0}. In ternary aquaous
solutions crossover to multicritical behavior was reported
\cite{anisimov:00:0}, and a tricritical, rather than a critical point 
was indeed observed in simulations~\cite{dickman:99:0,panag:99:0,diehl:03:0} 
and predicted theoretically~
\cite{hoye:97:0,stell:99:0,dickman:99:0,ciach:00:0,ciach:01:1,kobelev:02:0,brognara:02:0} in the RPM with 
the locations of the ions restricted to the sites of the simple-cubic
lattice. The first firm prediction of tricriticaliy for that system
appeared in Ref.\cite{hoye:97:0}, where a strong argument based on
symmetry considerations was given.

 The puzzling contradictions described above motivated careful
 theoretical, simulation and experimental studies of criticality in
 ionic sytems. In simple words, the Coulombic forces are long-range,
 and as such might lead to the mean-field (MF) criticality, as noted
 by Fisher~\cite{fisher:94:0}. On the other hand, screening effects
 lead to short-range effective forces, and standard fluid criticality
 might be expected
\cite{fisher:94:0,stell:95:0}. One of the important issues is the
range of the charge-density correlations in the critical region,
i.e. beyond the infinite dilution
regime~\cite{fisher:94:0,stell:95:0}.

 Subsequent precise experiments
by Schr\"oer, Wiegand and others
\cite{wiegand:94:0,wiegand:97:0,wiegand:98:0,wiegand:98:1,
kleemeier:99:0,wagner:02:0,wagner:03:0,wagner:04:0} showed that by
careful identification and elimination of various additional effects
(chemical instability, background scattering) Ising criticality is
obtaind in all the systems for which MF criticality or tricriticalty,
or a delayed crossover were reported previously. The authors of
Ref.\cite{anisimov:00:0} repeated the experiments and found that the
previous observations were associated with long-living metastable
states. After the equilibriation time (days) the Ising criticality was
observed~\cite{kostko:04:0}. At present no RPM-like experimental ionic
system deviates from the Ising criticality.  Recent simulations~
\cite{orkoulas:99:0,yan:99:0,luijten:01:0,caillol:02:0,luijten:02:0,kim:03:0}
 also strongly support the Ising criticality.

On the theoretical side the earlier arguments by
Stell~\cite{stell:92:0,stell:95:0}, and the field-theoretic
determination of the universality class, briefly sketched in
Refs.~\cite{ciach:00:0,ciach:02:0} are still considered as not fully
satisfactory. Since exact analytical calculations are impossible, as
is also the case for the three-dimensional Ising model, there is a
demand for a theory analogous to the renormalization-group (RG) theory in
simple fluids~\cite{amit:84:0,zinn-justin:89:0}. 
The latter  is restricted to short-range interactions, and is
not directly applicable to the Coulombic forces.

In this work the theory introduced in Ref.\cite{ciach:00:0} is further
developed, and exact derivations as well as technical details are
explained on a formal basis. We start with the mesoscopic description
first proposed in Ref.\cite{ciach:00:0} and described in more detail
in Refs.\cite{ciach:05:0,ciach:05:2}. In the coarse-grained description 
local distributions of
the ionic species occur with a probability proportional to the
Boltzmann factor $\exp (-\beta\Omega^{MF}[\phi({\bf x})\rho({\bf
x})])$. Here $\Omega^{MF}[\phi({\bf x})\rho({\bf x})]$ is the grand
potential in the system with the charge- and number densities
constrained to be $\phi({\bf x})$ and $\rho({\bf x})$ in an
infinitesimal ('mesoscopic') volume $d{\bf x}$ centered at the space
position $ {\bf x}$. It seems natural to reduce the two-field theory
to a one-field theory depending only on the number-density of ions,
which should be a proper order parameter for the gas-liquid type
transition. However, in the RPM only charges interact, and it is not
possible to perform such a reduction in a simple way.  On the other
hand, because there are no interactions between the masses in the RPM,
the field $\rho({\bf x})$ can be integrated out easily. In sec.3 we
reduce the two-field theory to the one-field theory of the field
$\phi({\bf x})$, with the Boltzmann factor $\propto\exp (-\beta {\cal
H}_{eff}[\phi])$. The particular feature of the one-field theory is
that the second functional derivative of ${\cal H}_{eff}[\phi]$
assumes a minimum for the wavenumber $k_b\ne 0$ in Fourier
representation, and the higher-order terms $\propto \int_{\bf x}{\cal
A}_{2n}\phi^{2n}({\bf x})$ have the property that for low valuse of
the average number density $ {\cal A}_4<0$ and $ {\cal A}_{2n}>0$ for $n>2$.
We find exact expressions for the number-density correlation functions
in terms of the higher-order correlation functions for $\phi({\bf x})$
in the one-field theory.

In order to determine the scaling properties of the connected
number-density correlation functions we consider the corresponding
Feynman diagrams in the (formal) perturbation expansion in $ {\cal
A}_{2n}$. In sec.4 we study the two-point charge-density correlation
function $G_{\phi\phi}$. In the Gaussian approximation $\tilde
G^0_{\phi\phi}(k)$ is singular for the wavenumber $k_b\ne 0$. The
singularity is removed beyond the Gaussian approximation by following
the Brazovskii theory~\cite{brazovskii:75:0}. We segregate all Feynman
diagrams for any given connected number-density correlation function
into disjoint sets according to a prescription described in sec.5. By
summing all diagrams in each set we obtain a finite expression, which
can be represented by a skeleton diagram with lines representing the
charge-density correlation function. We call such a series of diagrams
'a secondary diagram'. In this way we cure the artificial singularity
of individual diagrams following from the singularity of $\tilde
G^0_{\phi\phi}(k_b)$.

The secondary diagrams are again segregated into disjoint sets,
according to their topological properties, as explained in sec.  5. A
series of the secondary diagrams belonging to a particular set can be
represented by a hyperdiagram. The hyperdiagrams are associated with
the same expressions for $k\to 0$ as the corresponding diagrams in a
model system with the Hamiltonian belonging to the Ising universality
class, which contains also irrelevant operators. We could repeat the
whole procedure of regularization of the hyperdiagrams,
renormalization of the coupling constants and solve the flow equations
describing the evolution of the renormalized coupling constants under
changing the length scale, but this has already been
done~\cite{amit:84:0,zinn-justin:89:0} for the model Hamiltonian
yielding the same expressions for the vertex functions in the (formal)
perturbation expansion.  In sec.5 we also describe corrections to
scaling specific for the RPM. The last section contains a short summary.

\section{Background}
\subsection{Coarse-grained description for the RPM. }
In the field-theoretic, coarse-grained approach we consider
 local densities of the ionic species in mesoscopic regions $d{\bf x}$ 
around each point ${\bf x}$,
$\rho_{\alpha}({\bf x})$, where $\alpha=+,-$. For fixed temperature
and chemical potential the probability density that the local
densities assume a particular form $\rho_{+}({\bf x})$, $\rho_{-}({\bf
r})$ is given by \cite{ciach:05:0,ciach:05:2}
\begin{equation}
\label{Boltz}
p[\rho_{+}^*({\bf x}),\rho_{-}^*({\bf x})]=\Xi^{-1}
\exp(-\beta \Omega^{MF}[\rho_+^*,\rho_-^*]),
\end{equation}
where
\begin{equation}
\label{Xi}
\Xi= \int D\rho^*_+\int D\rho^*_- 
\exp(-\beta \Omega^{MF}[\rho_+^*,\rho_-^*] ),
\end{equation}
$\beta=1/kT$ with $k$ the Boltzmann constant and $T$ temperature,
and we introduced dimensionless densities
\begin{equation}
\rho^*_{\alpha}=\rho_{\alpha}\sigma_{+-}^3,
\end{equation}
where  $\sigma_{+-}=(\sigma_{+}+\sigma_{-})/2$ is the sum of radii.
In the above $ \Omega^{MF}[\rho_+^*,\rho_-^*]$
is the grand potential in the system where the local concentrations of
the two ionic species are constrained to be  $\rho^*_{+}({\bf
r})$, $\rho^*_{-}({\bf x})$,
\begin{equation}
\label{Omega}
\Omega^{MF}[\rho^*_+,\rho^*_-] =
F_h[\rho^*_+,\rho^*_-] + U^{PM}[\rho^*_+,\rho^*_-] 
-\mu_{\alpha}\int_{\bf x} \rho^*_{\alpha}({\bf x}),
\end{equation}
We use simplified notation $\int_{\bf x}\equiv \int d{\bf x}$
throghout the whole paper. In Eq.(\ref{Omega}) $F_h$ is the Helmholtz
free energy of the hard-core reference system, and we shall limit
ourselves to the local-density approximation
$F_h[\rho^*_+,\rho^*_-]=\int d{\bf x} f_h(\rho^*_+({\bf x}),
\rho^*_-({\bf x}))$. In principle, more accurate approximations for
  $F_h[\rho^*_+,\rho^*_-]$ could be adopted. 
In the local density approximation 
\begin{equation}
\label{fh}
\beta \frac{\partial^2f_h}{\partial\rho_{\alpha}^*\partial\rho_{\beta}^*}
=\frac{\delta^{Kr}_{\alpha,\beta}}{\rho^*_{\alpha}}-c_h(\rho^*),
\end{equation}
where the first term results from the ideal entropy of mixing, 
\begin{equation}
\rho^*=\rho^*_++\rho^*_-,
\end{equation}
and
$c_h(\rho^*)$
 is the volume integral over the hard-sphere
contribution to the Ornstein-Zernike direct correlation function.

 In this
work we focus on the RPM. In the RPM $\delta^2
\Omega^{MF}/\delta\rho^*_{\alpha}\delta\rho^*_{\beta}$ can be easily
diagonalized due to the symmetry of the interaction potentials, and
the eigenmodes have a natural physical interpretation. The first
eigenmode is the number-density deviation of ionic species from the
most probable value,
\begin{equation}
\eta({\bf x})=\rho^* ({\bf x})-\rho^*_0,
\end{equation}
 and the second mode is the local
 charge-density  in $e$-units ($e=e_+=|e_-|$), 
\begin{equation}
\phi({\bf x})=\rho^*_+({\bf x})-\rho^*_-({\bf x}).
\end{equation}
 The electrostatic energy is 
%
\begin{equation}
\label{UPM}
\beta U^{RPM}[\phi]=
\frac{\beta^*}{2}\int_{{\bf x}_1}\int_{{\bf x}_2}  
\phi({\bf x}_1)V({\bf x}_1-{\bf x}_2)
\phi({\bf x}_2),
\end{equation}
\begin{equation}
\label{beta*}
\beta^*=\frac{1}{T^*}=\frac{\beta e^2}{D\sigma},
\end{equation}
\begin{equation}
\label{Vv}
  V({\bf x})=
\frac{\theta(x-\sigma_{\alpha\beta})}{x},
\end{equation}
and $x=|{\bf x}|$. 
In the following,  distances will  be measured  in
$\sigma_{+-}=\sigma$ units.
In Eq. (\ref{UPM}) the contributions to the electrostatic energy
 coming from overlapping hard spheres are excluded because of the form
 of the potential $V$ (see Eq.(\ref{Vv})).

 In the field theory introduced above the physical quantities are
 obtained by averaging over all fields $\rho_+^*$, $\rho_-^*$ (or over 
 $\phi$ and $\eta$ in the RPM) with the Boltzmann factor (\ref{Boltz}), and the
 grand potential $\Omega$ is
\begin{equation}
\label{grdef}
-\beta\Omega=\log \Xi.
\end{equation}
 For small amplitudes of the fields $\phi$ and $\eta$
the functional (\ref{Omega}) can be expanded about its
value $\Omega^{MF}_0$ at the minimum,
\begin{equation}
\label{OmF}
\Delta \Omega^{MF}=\Omega^{MF}- \Omega^{MF}_0=\Omega_2+\Omega_{int}.
\end{equation}
Here $\Omega_2$ denotes the Gaussian part of the functional. 
  In terms of the eigenmodes the Gaussian part of the functional
 (\ref{OmF})
  assumes the form
\begin{equation}
\label{Gau}
\beta\Omega_2={1\over 2}\int_{\bf k}\Bigg[
\tilde C_{\phi\phi}^0({\bf k})\tilde \phi({\bf k})\tilde\phi(-{\bf k})
+\tilde C_{\eta\eta}^0({\bf k})\tilde\eta({\bf k})
\tilde \eta(-{\bf k})\Bigg],
\end{equation}
where we simplify the notation for the integrals in the Fourier space 
$\int_ {\bf k}\equiv\int{d {\bf k}\over (2\pi)^3} $. By using (\ref{fh}) we obtain
\begin{equation}
\label{phitad}
\tilde C_{\phi\phi}^0({\bf k})=  \rho_0^{*-1}+\beta^* \tilde V({\bf k}),
\end{equation}
and
\begin{equation}
\label{Cetad}
\tilde C_{\eta\eta}^0({\bf k})= \gamma_{0,2}=
\beta {\partial^2f_h\over \partial\rho^{*2}}
\vert_{\phi=0,\rho^*=\rho^*_0} .
\end{equation}
In the continuum-space RPM the Fourier transform of the potential (\ref{Vv}) 
has the form
\begin{equation}
\label{vfrpm}
\tilde V(k)=\frac{4\pi\cos k}{k^2},
\end{equation}
 where $k=|{\bf k}|$ is in $\sigma^{-1}$ units. The
 wavenumber corresponding to the minimum of the potential $\tilde V(k) $ 
 is $ k_b\ne 0$
\cite{ciach:00:0}.  The explicit form of $\gamma_{0,2}$ depends on the
approximation for the hard-sphere reference-system; in any case, in
the absence of short-range attractive forces it is a positive function
of $\rho^*_0$. Hence, $\eta$ is a noncritical field. On the other hand, $
\tilde C_{\phi\phi}^0( k)$ can vanish for $k\ne 0$.
 The boundary of stability of $\Delta\Omega^{MF}$ in the $(\rho^*,T^*)$ 
phase space is given by 
\begin{equation}
\label{bound_stab}
 \tilde C_{\phi\phi}^0(k_b)=0.
 \end{equation}
The critical fluctuations are thus $\tilde \phi({\bf k}_b)$. 

The $\Omega_{int}$ has the expansion
\begin{eqnarray}
\label{Om_exp}
\beta\Omega_{int} [\phi,\eta]=
\sum_{m,n}^{\prime}
 \int_{{\bf x}}
\frac{\gamma_{2m,n}}
{(2m)!n!}\phi({\bf x})^{2m}
\eta({\bf x})^n.
\end{eqnarray} 
where $\sum_{m,n}^{'}$ denotes the summation with $2m+n\ge
3$, and $\gamma_{2m,n}$ denotes the
appropriate derivative of $f_h$ at $\phi=0$ and $\eta=0$.
\subsection{Ising Universality Class and Scaling}
The universality class of a given system is associated with a
particular scaling form of the singular part of the thermodynamic
potential in the critical region, or, equivalently, by the scaling
properties of the large-distance part of the connected correlation
functions. In particular, the Ising universality class is represented
by the standard $\varphi^4$ theory with the Hamiltonian
\begin{eqnarray}
\label{phi44}
{\cal H}^I[\varphi]=\frac{1}{2}
\int_{\bf k}(t_0+k^2)\tilde\varphi({\bf k})
\tilde\varphi(-{\bf k})+\int_{\bf x}\Big(-H_0\varphi({\bf x})+
\frac{u_0}{4!}\varphi^4({\bf x})\Big).
\end{eqnarray}
The (renormalized) connected $N$-point correlation function for the
field $\varphi$ scales in the real-space representation according to
\cite{amit:84:0}
\begin{eqnarray}
\label{Issc}
G^R_N({\bf x}_1,...,{\bf x}_N;t,u,H)=
t^{N\beta}G^R_{N}({\bf x}_1t^{\nu},...,{\bf x}_Nt^{\nu};
1,u^*,Ht^{-\Delta}),
\end{eqnarray}
where $t$, $H$ and $u$ are the renormalized coupling constants
corresponding to the bare couplings $t_0$, $H_0$ and $u_0$
respectively, and $u^*$ is the fixed point of the RG flow of $\bar
u(\ell)$ upon rescaling the length unit, $x\to x\ell$ for $\ell\to 0$,
where $\bar u(\ell=1)=u$~\cite{amit:84:0}.  In the case of a magnet
$t$ and $H$ correspond to reduced temperature and magnetic field
respectively.  The two independent
critical exponents $\beta$ and $\Delta$ are related to the other
exponents via scaling relations
\cite{amit:84:0}. In particular,
$\beta=\nu(d-2+\eta)/2$ and $\Delta=\nu(d+2-\eta)/2$. 

Our purpose here is to determine the universality class of the RPM
critical point. Thus, we have to study scaling properties of the
correlation functions for the number-density field $\eta({\bf x})$
near the critical point of the phase separation, i.e. in the phase
space region where $\langle\phi\rangle =0$. It is the correlation
function for large separations, which may lead to divergent
susceptibility and obeys scaling. In determining the universality
class of a critical point one can separate the correlation function
into a long- and a short-distance parts, and neglect the latter. In the
next section we construct an effective one-field theory which allows
for a determination of nonlocal parts of the number-density 
correlation functions. 
\section{Effective field theory}
In this section we derive an exact effective Hamiltonian
 ${\cal H}_{eff}[\phi]$, such that the grand potential is given by
$\Omega=-kT\log \int D\phi \exp(-\beta{\cal H}_{eff}[\phi])$. Next we derive 
exact expressions for the nonlocal parts of the
number-density correlation functions in terms of higher-order
correlations for the field $\phi$, with the Boltzmann factor
 $\exp(-\beta{\cal H}_{eff}[\phi])$. We also introduce generating 
functionals for the relevant  correlation and vertex functions.
\subsection{Derivation of the effective functional}
The functional $\Delta \Omega^{MF}[\phi,\eta]$ (Eqs.(\ref{OmF}) and
(\ref{Omega})) can be split into two parts,
\begin{equation}
\label{split}
\Delta \Omega^{MF}[\phi,\eta]=\Omega_{\phi}[\phi]+
\Omega_{\eta}[\phi,\eta]
\end{equation}
where
\begin{equation}
\label{op}
\beta\Omega_{\phi}[\phi]=
\frac{1}{2}\langle\phi|C^0_{\phi\phi}|\phi\rangle + 
\sum_{m=2}\frac{\gamma_{2m,0}}{(2m)!}\langle 1|\phi^{2m}\rangle,
\end{equation}
\begin{equation}
\label{oep}
\beta\Omega_{\eta}[\phi,\eta]=
\frac{1}{2}\langle\eta|C^0_{\eta\eta}|\eta\rangle+
\sum_{m,n}^{\prime\prime}
\frac{\gamma_{2m,n}}{(2m)!n!}\langle
\phi^{2m}|\eta^n\rangle,
\end{equation}
and where $\sum_{m,n}^{\prime\prime}$ indicates a summation with $n\ge 1$
and $2m+n\ge 3$. In the real space representation $\phi$ and $\eta$
are real functions. For convenience we use the Dirac brac-kets
notation for the scalar product
\begin{equation}
\label{scpr}
\langle f|g\rangle\equiv\int_{\bf x} f^*({\bf x})
 g({\bf x})= \int_{\bf k} \tilde f({\bf k})
 \tilde g(-{\bf k}),
\end{equation}
and  $|A|f\rangle$ denotes the operator $A$ acting on the state
$|f\rangle$. In particular, when in real-space representation $A$ has a 
functional
 form $A({\bf
x},{\bf x}')$, then
\begin{equation}
\label{acop}
|A|f\rangle\equiv \int_{{\bf x}'}A({\bf x},{\bf x}')f({\bf x}').
\end{equation}

We introduce a functional of two external fields $J({\bf x})$ and 
$v({\bf x})$ by
\begin{equation}
\label{fuwf}
\Delta\Xi[J,v]=\int D\phi e^{\langle J|\phi\rangle}
e^{-\beta\Omega_{\phi}[\phi]}
\Xi_{\eta}[\phi,v]
\end{equation}
where
\begin{eqnarray}
\label{Xiet}
\Xi_{\eta}[\phi,v]=\int D\eta 
e^{\langle v|\eta\rangle}e^{-\beta\Omega_{\eta}[\phi,\eta]}.
\end{eqnarray}
 Because in the pure RPM (no other interactions included)
 $C_{\eta\eta}^0({\bf x}_1,{\bf x}_2)=C_{\eta\eta}^0\delta({\bf
 x}_1,{\bf x}_2)$ (see Eq.(\ref{Cetad})),  Eq.(\ref{oep})  can be rewriten in the form
\begin{eqnarray}
\label{omlocal}
\beta\Omega_{\eta}[\phi,\eta]=
\int_{\bf x}\beta\omega_{\eta}(\phi({\bf x}),\eta({\bf x})),
 \end{eqnarray}
and we immediately obtain
the exact result
\begin{eqnarray}
\label{lXiet}
\log \Xi_{\eta}[\phi,v]=\int_{\bf x}\omega(\phi({\bf x}),v({\bf x})),
\end{eqnarray}
where
\begin{eqnarray}
\label{lXiet1}
\omega(\phi({\bf x}),v({\bf x}))=
\log\Bigg[\int d(\eta({\bf x}))
e^{-\beta\omega_{\eta}(\phi({\bf x}),\eta({\bf x}))}
e^{v({\bf x})\eta({\bf x})}\Bigg]
.
\end{eqnarray}
The integrand  $\omega(\phi({\bf x}),0)$ in Eq. (\ref{lXiet})
 can be expanded in a Taylor 
series with respect to $\phi({\bf x})$ and we obtain the formal 
expression
\begin{equation}
\label{lXe}
\omega(\phi({\bf x}),0)=
\sum_{m=0}^{\infty}
\frac{f_{2m}(\rho_0^*)}{(2m)!}\phi^{2m}({\bf x}),
\end{equation}
where the explicit forms of $f_{2m}(\rho_0^*)$ depend on the form of
the Helmholtz free-energy density  of the hard-sphere reference system.  By
substituting the above form of $\Xi_{\eta}$ into Eq.(\ref{fuwf}) we
obtain
\begin{equation}
\label{Ffuu}
\Delta\Xi[J,0]=Z_{\eta}\Xi_0[J],\hskip1cm 
\Xi_0[J]=\int D\phi e^{\langle J|\phi\rangle}
e^{-\beta{\cal H}_{eff}[\phi]}
\end{equation}
where $\log Z_{\eta}=\int_{\bf x} f_0(\rho_0^*)$ ($\log \Delta\Xi[J,0]$
 is an extensive quantity), 
\begin{equation}
\label{hef}
\beta{\cal H}_{eff}[\phi]=
\frac{1}{2}\langle\phi|{\cal C}_{\phi\phi}^0|\phi\rangle
+\sum_{m=2}^{\infty}\frac{{\cal A}_{2m}}{(2m)!}
\int_{\bf x}\phi^{2m}({\bf x}),
\end{equation}
\begin{equation}
\label{calC}
{\cal C}_{\phi\phi}^0({\bf x},{\bf x}')=C_{\phi\phi}^0({\bf x},{\bf x}')+ 
f_2(\rho_0^*)\delta({\bf x}-{\bf x}')
\end{equation}
and 
\begin{equation}
\label{calAn}
{\cal A}_{2m}=\gamma_{2m,0}+f_{2m}(\rho_0^*).
\end{equation}
Eqs.(\ref{Ffuu})-(\ref{calAn}) define the effective functional.

Eq.(\ref{Xiet}) can be written in a more general form, 
valid also beyond the RPM (i.e. for nonlocal $C^0_{\eta\eta}$), 
 \cite{zinn-justin:89:0,amit:84:0}
\begin{eqnarray}
\label{Xiet2}
\Xi_{\eta}[\phi,v]=Z_{\eta}^0\cdot \exp\Bigg[-\sum_{m,n}^{\prime\prime}
\frac{\gamma_{2m,n}}{(2m)!n!}
\int_{\bf
x}\phi^{2m}({\bf x}) \frac{\delta^n}{\delta v({\bf   x})^n}\Bigg]
e^{\frac{1}{2}\langle v|G_{\eta\eta}^0|v\rangle}.
\end{eqnarray}
In the above the constant is $Z^0_{\eta}=\int D fe^{-\frac{1}{2}
\langle f|C_{\eta\eta}^0|f\rangle}$, and 
\begin{eqnarray}
\label{zea}
Z_{\eta}=
Z_{\eta}^0\exp\Bigg[-\sum_{n=3}^{\infty}\frac{\gamma_{0,n}}{n!}
\int_{\bf x}\frac{\delta^n}{\delta v({\bf x})^n}
\Bigg]e^{\frac{1}{2}\langle v|G^0_{\eta\eta}|v\rangle}|_{v=0}.
 \end{eqnarray}
  Nonvanishing contributions to
$\Xi_{\eta}[\phi,0]$ result from an even number of differentiations
 in Eq.(\ref{Xiet2})
with respect to $v$. In the case of the RPM 
$\langle f|G_{\eta\eta}^0|f\rangle=G_{\eta\eta}^0\langle f|f\rangle$, 
and a double differentiation with respect to $v$ 
gives a factor
$G_{\eta\eta}^0=1/C_{\eta\eta}^0$. The Wick theorem allows to
represent $\Xi_{\eta}[\phi,0]$ as a sum of all vacuum Feynman diagrams 
(no
external points) with vertices $\gamma_{2m,n}\phi^{2m}({\bf x})$ such 
that
$n\ge 1$ and $2m+n\ge 3$, all at the same point ${\bf x}$. From the 
vertex $\gamma_{2m,n}\phi^{2m}({\bf x})$
there emanate $n$ $\eta$-lines, and all $\eta$-lines have to be
paired. Lines connecting different vertices as well as loops represent
$G_{\eta\eta}^0$. Finally, each diagram is integrated over ${\bf x}$.

$\log\Xi_{\eta}[\phi,0]=\int_{\bf x}\omega(\phi({\bf x}),0) $ 
is represented
by all {\it connected} vacuum diagrams described above
\cite{amit:84:0,zinn-justin:89:0}. $\omega(\phi({\bf x}),0)$
 is represented 
by the same
diagrams, except that there is no integration over ${\bf x}$.  The
term proportional to $\phi^{2m}({\bf x})$ in the expansion of 
$\omega(\phi({\bf x}),0)$ in Eq.(\ref{lXe}) 
is represented by the sum of all connected diagrams, such that each
diagram in the sum contains a certain number $N$ of vertices
$\gamma_{2m_j,n_j}\phi^{2m_j}({\bf x})$, and $\sum_{j=1}^N m_j=m$. In
particular, $f_2
\phi^{2}({\bf x})$ is represented by a sum of all connected vacuum
diagrams, where each diagram in the sum contains an arbitrary number
of vertices $\gamma_{0,n}$, and a single vertex
$\gamma_{2,k}\phi^{2}({\bf x})$, where $k$ is arbitrary.  There are no
zero-loop contributions to  $f_2
\phi^{2}({\bf x})$, and the
one-loop contributions are shown in Fig.1. 
\begin{figure}
\includegraphics[scale=0.5]{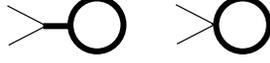}
\caption{One-loop diagrams contributing to $f_2
\phi^{2}({\bf x})$. Thick lines and loops represent $G^0_{\eta\eta}$ and thin 
lines represent $\phi$. Vertices with $2m$ thin lines and $n$ thick lines 
represent $\gamma_{2m,n}$.}
\end{figure}
\begin{figure}
\includegraphics[scale=0.5]{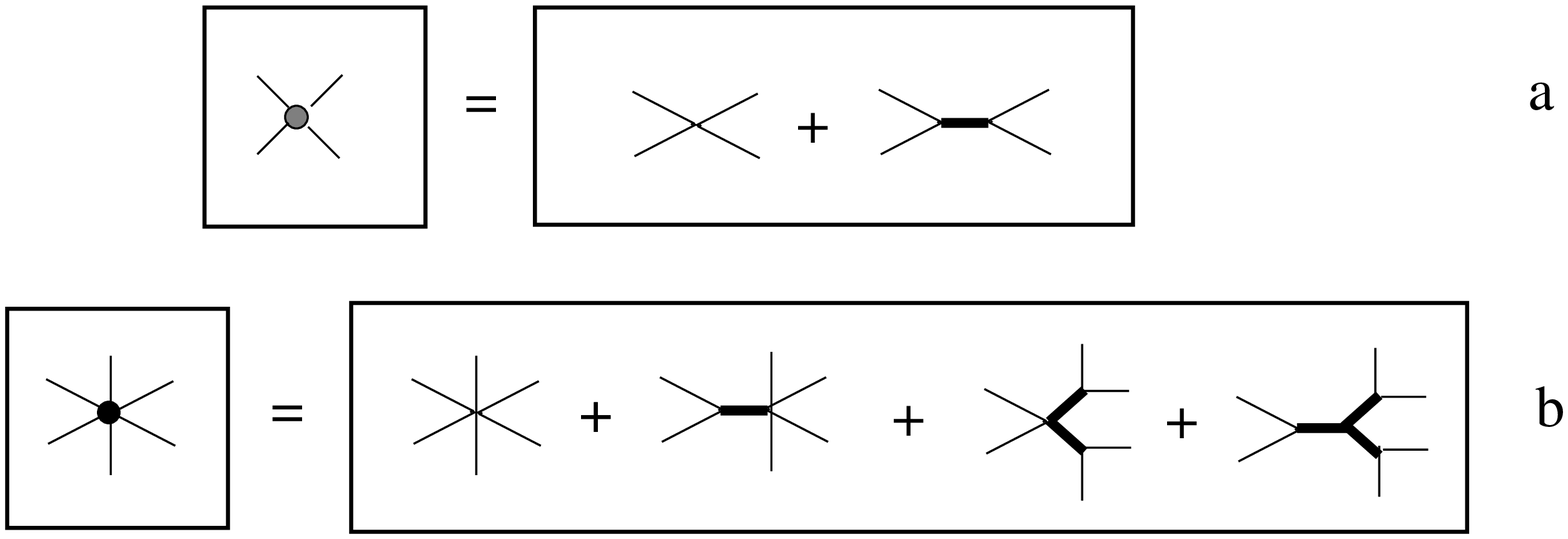}
\caption{Zero-loop diagrams contributing to ${\cal A}_4\phi^4({\bf x})$ 
(a) and  
${\cal A}_6\phi^6({\bf x})$ (b).
 Thick lines  represent $G^0_{\eta\eta}$ and thin 
lines represent $\phi$. Vertices with $2m$ thin lines and $n$ thick lines 
represent $\gamma_{2m,n}$.           
 }
\end{figure}
\begin{figure}
\includegraphics[scale=0.5]{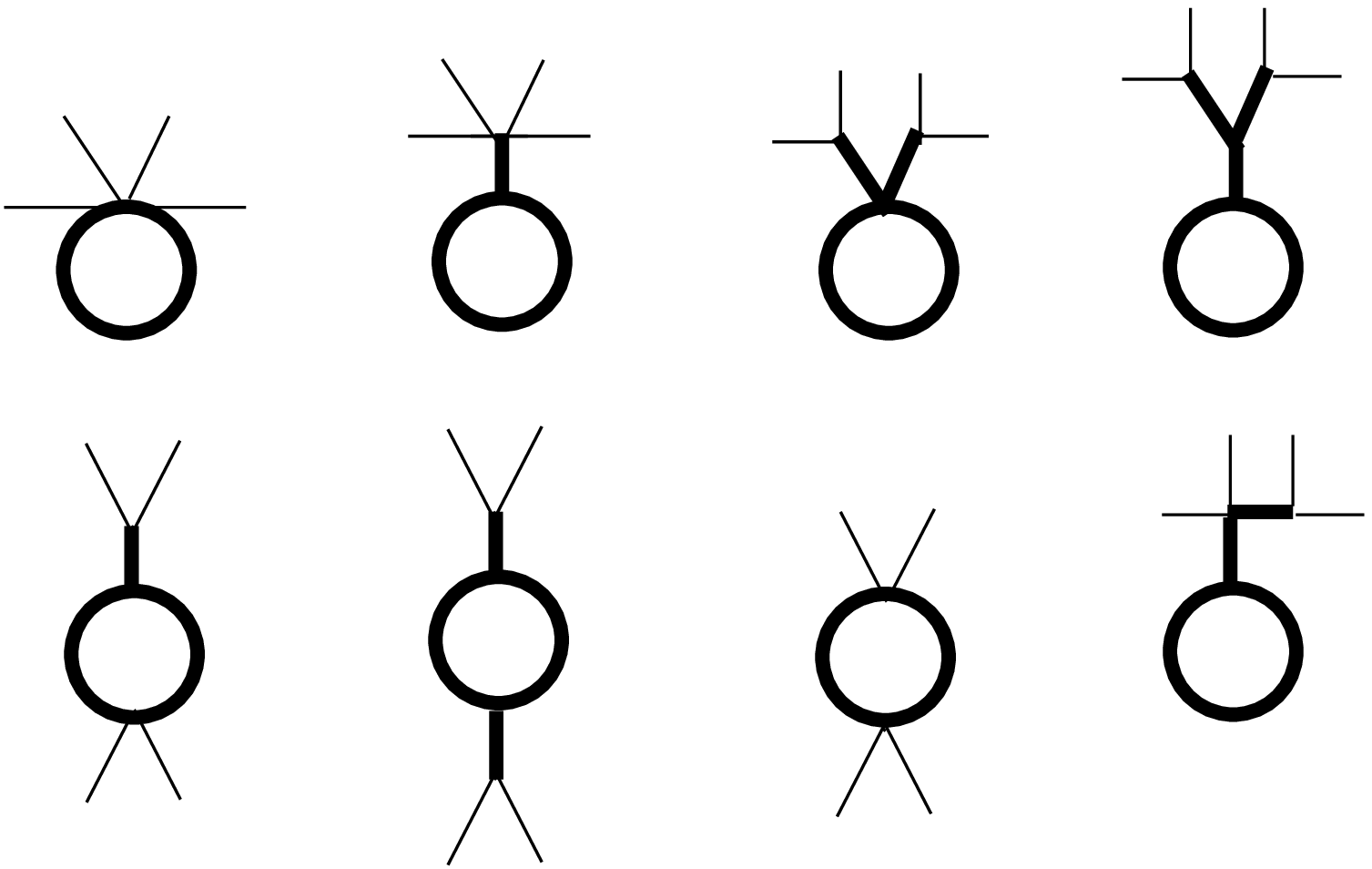}
\caption{One-loop diagrams contributing to  ${\cal A}_4\phi^4({\bf x})$.
 Thick lines and loops represent $G^0_{\eta\eta}$ and thin 
lines represent $\phi$. Vertices with $2m$ thin lines and $n$ thick
lines represent $\gamma_{2m,n}$.}
\end{figure}
 The zero-loop diagrams contributing to ${\cal A}_{4}\phi^{4}({\bf
 x})$ and ${\cal A}_{6}\phi^{6}({\bf x})$ are shown in Fig.2, and one
 loop diagrams contributing to ${\cal A}_4\phi^{4}({\bf x})$ are shown
 in Fig.3.  In these diagrams each thin line emanating from the vertex
 at ${\bf x}$ represents $\phi({\bf x})$,  thick lines represent
 $G_{\eta\eta}^0$, and the diagrams are multiplied by appropriate
 symmetry factors \cite{amit:84:0,zinn-justin:89:0}.
 ${\cal A}_{2m}$ is represented by a sum of the corresponding
 diagrams with amputated $\phi({\bf x})$-lines.

The above perturbation approach is convenient for a development of
approximate theories.
Let us compare the above exact theory with the weighted-field (WF) 
approximation developed in  Refs.\cite{ciach:00:0,ciach:04:1}. The WF 
approximation is of the mean-field type with respect to the field 
$\eta$, because the effective functional is obtained by minimizing 
$\Delta\Omega^{MF}$ with respect to $\eta$ for a given field $\phi$.
 ${\cal H}_{eff}[\phi]$ given
in Eq.(\ref{hef}) has the same functional form as in the WF
theory. Moreover, in the zero-loop approximation the term $f_2$
vanishes, and the coefficients ${\cal A}_{2m}$ reduce to those
obtained previously in Ref.
\cite{ciach:04:1}.  Thus, in the zero-loop approximation the above exact
effective theory reduces to the  WF approximation, as expected.

In the effective theory derived above we can rewrite $\Xi_0[J]$
in the form
\begin{equation}
\label{kuku}
\Xi_0[J]=Z_{\phi}\Xi_1[J],\hskip1cm
\Xi_1[J]= e^{-\sum_{m=2}\frac{{\cal A}_{2m}}{2m!}
\int_{\bf x}\frac{\delta^{2m}}{\delta  J({\bf x})^{2m} }}
e^{\frac{1}{2}\langle J| G_{\phi\phi}^0|J\rangle},
\end{equation}
where $Z_{\phi}=\int D f e^{-\frac{1}{2}\langle f|{\cal
C}_{\phi\phi}^0|f\rangle}$, and the charge-density correlation
function in the Gaussian approximation, $ G_{\phi\phi}^0$, is inverse
to ${\cal C}_{\phi\phi}^{0}$,
\begin{equation}
\label{gpi}
\tilde G_{\phi\phi}^0(k)=1/\tilde{\cal C}_{\phi\phi}^{0}(k).
\end{equation}
  By applying the Wick theorem we can represent $\Xi_1[0]$ by vacuum
  diagrams with 'hypervertices' ${\cal A}_{2m}$ from which there
  emanate $2m$ $\phi$-lines. All lines are paired, and the line
  connecting vertices at ${\bf x}_1$ and ${\bf x}_2$ represents $
  G_{\phi\phi}^0({\bf x}_1,{\bf x}_2)$. The name 'hypervertex' used in
  Ref.\cite{ciach:04:1} comes from the form of ${\cal A}_{2m}$ in the
  original, two-field theory (see Eq.(\ref{calAn}) and Fig.2), but in
  the effective one-field theory ${\cal A}_{2m}$ are just usual
  vertices. For simplicity, in the rest of this paper we shall use the
  name 'vertex' for ${\cal A}_{2m}$.
\subsection{Stability of the functional ${\cal H}_{eff}$}
The effective functional of the form (\ref{hef}) was already studied
in the WF theory \cite{ciach:04:1,ciach:05:0}, which is equivalent to
the zero-loop approximation for $f_{2}$ and ${\cal A}_{2n}$.
 Beyond the WF approximation the Gaussian correlation function $
\tilde G_{\phi\phi}^0({\bf k})$ diverges for the same wavevector ${\bf
k}={\bf k}_b\ne {\bf 0}$ as found in Ref.\cite{ciach:00:0}, but along
the spinodal line
\begin{equation}
\label{sp}
T^*_b=-\rho_0^*\frac{\tilde V({\bf k}_b)}{1+\rho_0^*f_2(\rho_0^*)} ,
\end{equation}
 which is shifted compared to that found in the WF theory, 
where $f_2=0$ (note that
 $\tilde V({\bf k}_b)<0$). We assume that the denominator is positive
 for the relevant range of $\rho_0^*$. For particular forms of $f_h$
 this assumption remains to be verified. Divergent $ \tilde
 G_{\phi\phi}^0({\bf k}_b)$ for ${\bf k}_b\ne {\bf 0}$ indicates that
 our theory belongs to the class of the Brazovskii field theories 
\cite{brazovskii:75:0}.

Another important property of our WF theory is the fact that for low
densities, $\rho_0^*<\rho_{tc}^*$, the vertex ${\cal A}_4$
becomes negative, and ${\cal A}_{2n}>0$ for $n>2$
\cite{ciach:00:0,ciach:04:1}. This property leads to a tricritical
point on the sc lattice \cite{ciach:03:0,ciach:04:1}, and to a
gas-liquid type  instability in continuum and some other lattice
systems \cite{ciach:03:0,ciach:04:0}, in agreement with simulation
results \cite{panag:99:0,diehl:03:0,diehl:05:0}. Based on the WF and
the simulation results we focus here on the general field theory
having the property that ${\cal A}_4(\rho_0^*)$ becomes negative for
sufficiently low densities, with ${\cal A}_6>0$
\cite{ciach:05:0,ciach:04:1}. Within the exact theory
developed above this assumption remains to be verified for particular
forms of the reference system. Negative values of ${\cal A}_4$ for low
densities bring our field theory in the corresponding part of the
phase diagram beyond the class of the Brazovskii-type field theories,
where ${\cal A}_4>0$.
\subsection{Correlation functions in the effective field theory}
The correlation functions for the field $\phi$ can be obtained in a
standard way from their generating functional $\Delta\Xi[J,v]$ given in
 Eq.(\ref{fuwf}),
\begin{equation}
\label{munu}
\langle\langle\phi^{k_1}({\bf x}_1) ... \phi^{k_n}({\bf x}_n)\rangle\rangle=
\frac{1}{\Delta\Xi}\frac{\delta^{k_1+...+k_n}\Delta\Xi}
{\delta J({\bf x}_1)^{k_1}...\delta J({\bf x}_n)^{k_n}
}\vert_{J=0,v=0},
\end{equation}
and the connected correlation functions are given by 
\cite{amit:84:0,zinn-justin:89:0}
\begin{equation}
\label{munucon}
\langle\langle\phi^{k_1}({\bf x}_1) ... \phi^{k_n}({\bf x}_n)\rangle\rangle^{con}=
\frac{\delta^{k_1+...+k_n}\log\Delta\Xi}
{\delta J({\bf x}_1)^{k_1}...\delta J({\bf x}_n)^{k_n}}
\vert_{J=0,v=0}.
\end{equation}
$\langle\langle ...\rangle\rangle$ denotes the average  in the
original two-field theory.  
From (\ref{fuwf}) and  (\ref{Ffuu}) we obtain 
\begin{equation}
\langle\langle\phi^{k_1}({\bf x}_1) ... \phi^{k_n}({\bf x}_n)\rangle\rangle=
\langle\phi^{k_1}({\bf x}_1) ... \phi^{k_n}({\bf x}_n)\rangle,
\end{equation}
 where $\langle ... \rangle$ denotes averaging in the effective one-field
theory.
The connected correlation function
$\langle\phi^{k_1}({\bf x}_1) ... \phi^{k_n}({\bf x}_n)\rangle^{con}$
is represented by a sum of all connected diagrams containing external
points $ {\bf x}_1,... {\bf x}_n$ from which there emanate
$k_1,...,k_n$ $\phi$-lines respectively, and vertices ${\cal
A}_{2m}$ from which there emanate $2m$ $\phi$-lines. All lines are
paired. A line connecting a vertex or an external point at ${\bf x}_1$
with a vertex or an external point at ${\bf x}_2$ represents $
G_{\phi\phi}^0({\bf x}_1,{\bf x}_2)$ \cite{amit:84:0,zinn-justin:89:0}.

Let us consider the number-density correlation functions 
defined by
\begin{equation}
\label{munue}
\langle\langle\eta({\bf x}_1) ... \eta({\bf x}_n)\rangle\rangle=
\frac{1}{\Delta\Xi}\frac{\delta^{n}\Delta\Xi}
{\delta v({\bf x}_1)...\delta v({\bf x}_n)}\vert_{J=0,v=0}.
\end{equation}
 From (\ref{fuwf}) we have 
\begin{equation}
\label{munue1}
\langle\langle\eta({\bf x}_1) ... \eta({\bf x}_n)\rangle\rangle=
\frac{1}{\Delta\Xi}
\int D\phi e^{-\beta \Omega_{\phi}[\phi]}
D_n[{\bf x}_1,...,{\bf x}_n|\phi]
\end{equation}
where
\begin{equation}
\label{Dndef}
D_n[{\bf x}_1,...,{\bf x}_n|\phi]=
\frac{\delta^n\Xi_{\eta}[\phi,v]}
{\delta v({\bf x}_1)... \delta v({\bf x}_n)}|_{v=0}.
\end{equation}
By using Eq.(\ref{Xiet2}) for local $C_{\eta\eta}^0$, in the case 
of ${\bf x}_1\ne {\bf x}_2\ne
...\ne {\bf x}_n$ we obtain
\begin{equation}
\label{Dn}
D_n[{\bf x}_1,...,{\bf x}_n|\phi]=Z^0_{\eta}\cdot G_{\eta\eta}^{0n}
\exp\Bigg[-\sum_{m,n}^{''}\int_{\bf x}
\frac{\gamma_{2m,n}}{(2m)!n!}\phi^{2m}({\bf x})
\frac{\delta^n}{\delta v({\bf x})^n}\Bigg] v({\bf x}_1)... v({\bf x}_n)
e^{\frac{1}{2}G_{\eta\eta}^0\langle v|v\rangle}|_{v=0}.
\end{equation}
When ${\bf x}_i={\bf x}_j$ for at least one pair of external points, 
there are additional contributions to 
$D_n[{\bf x}_1,...,{\bf x}_n|\phi]$ \cite{amit:84:0,zinn-justin:89:0}.
We shall 
focus only on  the case of ${\bf
x}_1\ne {\bf x}_2\ne ...\ne {\bf x}_n$. From the Wick theorem it
follows that in the case of the RPM each diagram contributing to $D_n$
consists of $n$ disjoint diagrams, each of them containing a single
external point, and of vacuum diagrams. Thus,
\begin{equation}
\label{Dn2}
D_n[{\bf x}_1,...,{\bf x}_n|\phi]=\Xi_{\eta}[\phi,0]
\prod_{i=1}^{n} \zeta({\bf x}_i),
\end{equation}
where 
\begin{equation}
\label{Dn3}
\zeta({\bf x}_i)=\sum_{m=0}\frac{a_{m}}{m!}\phi^{2m}({\bf x}_i),
\end{equation}
and $a_m$ are functions of $\rho_0^*$ depending on the form of the
free energy in the hard-sphere reference system.  Each term
$a_{m}\phi^{2m}({\bf x}_i)$ in Eq.(\ref{Dn3}) is represented by the sum
of all connected diagrams containing a single external point ${\bf
x}_i$, and an arbitrary number $N$ of vertices
$\gamma_{2m_j,n_j}\phi^{2m_j}({\bf x}_i)$ that satisfy $\sum_{j=1}^N
m_j=m$. Eqs. (\ref{Dn2}) and (\ref{Dn3}) are of course immediately
obtained by a direct differentiation of the exact formula
(\ref{lXiet}).

By substituting Eq.(\ref{Dn2}) into Eq.(\ref{munue})   we obtain
 for ${\bf x}_1\ne {\bf x}_2\ne ...\ne {\bf x}_n$ 
\begin{equation}
\label{etet}
\langle\langle\eta({\bf x}_1) ... \eta({\bf x}_n)\rangle\rangle=
\frac{1}{\Xi_0}\int D\phi 
e^{ -\beta{\cal H}_{eff}[\phi]}
 \zeta({\bf x}_1)...\zeta({\bf x}_n)= 
\langle\zeta({\bf x}_1) ... \zeta({\bf x}_n)\rangle.
\end{equation}
The
connected correlation function for the number-density fields can be
written in the form
\begin{equation}
\label{etetcon}
\langle\langle\eta({\bf x}_1) ... \eta({\bf x}_n)\rangle\rangle^{con}=
\langle\zeta({\bf x}_1) ... \zeta({\bf x}_n)\rangle^{con}=
\sum_{m_1=1}^{\infty}\frac{a_{m_1}}{m_1!}
...\sum_{m_n=1}^{\infty}\frac{a_{m_n}}{m_n!}
\langle \phi^{2m_1}({\bf x}_1)...\phi^{2m_n}({\bf x}_n)\rangle^{con}.
\end{equation}
  The above equation shows that for ${\bf x}_1\ne {\bf x}_2\ne ...\ne
  {\bf x}_n$ the connected number-density correlation functions are
  given in terms of higher-order charge-density correlation functions
  with the Boltzmann factor $\propto \exp\big(-\beta{\cal
  H}_{eff}[\phi]\big)$. We introduce for convenience an additional field
\begin{equation}
\label{psi}
\psi({\bf x})=a_1\phi^2({\bf x}),
\end{equation}
where
\begin{equation}
\label{a1}
a_1=G^0_{\eta\eta}\Big(-\frac{\gamma_{2,1}}{2}+...\Big),
\end{equation}
 and we rewrite Eq.(\ref{etetcon}) as follows
\begin{eqnarray}
\label{etetcon1}
\langle\zeta({\bf x}_1) ... \zeta({\bf x}_n)\rangle^{con}=
\sum_{m_1=1}^{\infty}\frac{a_{m_1}}{a_1^{m_1}m_1!}
...\sum_{m_n=1}^{\infty}\frac{a_{m_n}}{a_1^{m_n}m_n!}
\langle \psi^{m_1}({\bf x}_1)...\psi^{m_n}({\bf x}_n)\rangle^{con}=\\
\nonumber 
\langle\psi({\bf x}_1) ... \psi({\bf x}_n)\rangle^{con} + 
\frac{a_2}{a_1^2}\langle\psi^2({\bf x}_1) ... \psi({\bf x}_n)\rangle^{con}
 +... .
\end{eqnarray}
Note that $\langle\zeta({\bf x}_1) ... \zeta({\bf x}_n)\rangle^{con}$
is given in terms of $\langle\psi({\bf x}_1) ... \psi({\bf
x}_m)\rangle^{con}$ with $m\ge n$, such that for $m>n$ some of the external
points are identical, and the number of distinct external points is
$n$. $\langle\psi({\bf x}_1) ... \psi({\bf x}_n)\rangle^{con}$ is the
dominant contribution to $\langle\zeta({\bf x}_1) ... \zeta({\bf
x}_n)\rangle^{con}$ in the phase-space region where the
large-amplitude fluctuations are strongly damped by the Boltzmann
factor.

The phase separation, on which we focus in this work, is associated
with a long-distance behavior of the number-density correlation
function $\langle\langle\eta({\bf x}_1) \eta({\bf
x}_2)\rangle\rangle^{con}$. Contributions to $\langle\langle\eta({\bf
x}_1) \eta({\bf x}_2)\rangle\rangle^{con}$ which are proportional to $
\delta({\bf x}_1-{\bf x}_2)$ are irrelevant when the behavior of  
$\langle\langle\eta({\bf x}_1) \eta({\bf x}_2)\rangle\rangle^{con}$
for $|{\bf x}_1-{\bf x}_2|\to\infty$ is studied.  Because of that we
can approximate the exact correlation functions
$\langle\langle\eta({\bf x}_1) ... \eta({\bf
x}_n)\rangle\rangle^{con}$ by $ \langle\zeta({\bf x}_1) ... \zeta({\bf
x}_n)\rangle^{con}$. By doing so we neglect only the local parts of
the number-density correlations, because for ${\bf x}_1\ne {\bf
x}_2\ne ...\ne {\bf x}_n$ the above two functions are equal.  Note,
however that in the mesoscopic, or the coarse-grained theory we cannot
describe the structure for distances comparable to the size of ions.
Hence, the difference between $\langle\langle\eta({\bf x}_1)
... \eta({\bf x}_n)\rangle\rangle^{con}$ and $ \langle\zeta({\bf x}_1)
... \zeta({\bf x}_n)\rangle^{con}$ represents in fact the
microscopic-scale contribution to the correlation functions, which is
irrelevnt for the critical behavior. From now on we shall limit
ourselves to the one field theory and to the correlation functions
$\langle\psi({\bf x}_1) ... \psi({\bf x}_m)\rangle^{con}$, which
determine $\langle\zeta({\bf x}_1) ... \zeta({\bf x}_n)\rangle^{con}$
for $n\le m$.
\subsection{Vertex functions and the free-energy functional
 in the effective theory}
In the one-field theory we
 introduce an
effective generating functional for the connected correlation
 functions for the field  $\psi$ that is  defined in Eq.(\ref{psi}),
\begin{equation}
\label{genfun2}
\log \Xi_{eff}[w]=\log \int D\phi 
e^{\langle w|\psi\rangle}e^{-\beta {\cal H}_{eff}[\phi]}.
\end{equation}
The functional (\ref{genfun2}) generates in fact also the connected 
correlation
functions for the field $\zeta$, according to
Eq.(\ref{etetcon1}).
It is convenient to introduce a functional analogous to the
free-energy functional by a Legeandre transform of the above 
functional,
\begin{equation}
\label{gamfreff}
-\beta\Gamma_{eff}[\psi]=
\log \Xi_{eff}[w]-
\langle w|\psi\rangle,
\end{equation}
where in Eq.(\ref{gamfreff}) $\psi ({\bf x})=\delta\log\Xi_{eff}[w]/\delta w({\bf
x})$. The above functional can be expanded,
\begin{eqnarray}
\label{Gamma_exp3}
&\beta\Gamma_{eff} [\psi]= -
\langle F_{1}|\psi\rangle
+\frac{1}{2}\langle\psi|C_{\psi\psi}|
\psi\rangle  \nonumber \\
&+ \sum_{n>2}
\int_{{\bf x}_1}...\int_{{\bf x}_n}
\frac{F_{n}( {\bf x}_1,...,{\bf x}_n)}
{n!}
\psi({\bf x}_1)...\psi( {\bf x}_n).
\end{eqnarray} 
 $C_{\psi\psi}$ is inverse to $G_{\psi\psi}$, where we introduced
 the notation
\begin{equation}
\label{rr3}
G_{\psi\psi}(\Delta{\bf x})=
\langle\psi({\bf
 x})\psi({\bf x}+\Delta{\bf x})\rangle^{con}.
\end{equation}
 In diagrammatic expansion the vertex functions $F_{n}( {\bf
 x}_1,...,{\bf x}_n)$ are the one-particle irreducible (1PI) parts of
 the amputated $n$- point correlation functions, $\langle \psi({\bf
 x}_1)...\psi({\bf x}_n)\rangle^{con}_{amp}$, up to minus sign for
 $n\ge 3$. The 1PI diagrams cannot
 be split into two disjoint diagrams by cutting a single line. In the
  diagrams contributing to $\langle \psi({\bf
 x}_1)...\psi({\bf x}_n)\rangle^{con}_{amp}$ the lines connected with
 the external points are amputated,
\begin{eqnarray}
\langle \psi({\bf x}_1)...\psi({\bf x}_n)\rangle^{con}
=\int_{{\bf x}'}...\int_{{\bf x}^n}
\langle \psi({\bf x}')...\psi({\bf x}^n)\rangle^{con}_{amp}
 G_{\psi\psi}({\bf x}',{\bf x}_1)...
 G_{\psi\psi}({\bf x}^n,{\bf x}_n).
\end{eqnarray} 
The vertex functions $F_{n}( {\bf x}_1,...,{\bf x}_n)$ determine the
connected correlation functions $\langle \psi({\bf x}_1)...\psi({\bf
x}_n)\rangle^{con}$, which in turn determine $\langle \zeta({\bf
x}_1)...\zeta({\bf x}_m)\rangle^{con}$, i.e.  the nonlocal parts of
the connected  correlation functions for the field $\eta$. Thus, we have
reduced the two-field theory to the effective one-field theory.
\subsection{Strategy in determining  the universality class}
The universality class associated with the scaling properties of the
correlation functions $\langle \psi({\bf x}_1)...\psi({\bf
x}_n)\rangle^{con}$ could be inferred from the form of the functional
$H[\psi]$ such that
\begin{eqnarray}
\log \int D\psi e^{\langle w|\psi\rangle} e^{-\beta
H[\psi]}=\log \Xi_{eff}[w],
\end{eqnarray} 
where $\log \Xi_{eff}[w]$ is defined in Eq.(\ref{genfun2}). 
Determination of the exact form of $H[\psi]$ is not an easy
task. Note, however that we are not interested in the exact form of
the connected correlation functions, because scaling is obeyed only by
their long-distance parts. The latter are generated by the singular
part $\log\Xi^s_{eff}$ of the functional (\ref{genfun2}),
\begin{eqnarray}
\log
\Xi_{eff}[w]=\log \Xi^s_{eff}[w]+\log \Xi^r_{eff}[w],
\end{eqnarray} 
 where $ \log \Xi^r_{eff}$ denotes the regular part of the functional. In
 a similar way the functional (\ref{Gamma_exp3}) can be separated into
 the  singular and regular parts, 
$\Gamma_{eff}[\psi]= \Gamma^s_{eff}[\psi]+ \Gamma^r_{eff}[\psi]$,
obtained by the Legeandre transform
 of  $\log \Xi^s_{eff}[w]$ and $ \log \Xi^r_{eff}[w]$ respectively.

 Our strategy in determining the universality class consists of the
 following steps. First the connected number-density correlation functions 
and their 1PI parts are found within the perturbation expansion in the
 vertices ${\cal A}_{2m}$. Next we identify $\Gamma^s_{eff}[\psi]$, the
 singular part of $\Gamma_{eff}[\psi]$, and the 
 vertex functions generated by
 $\Gamma^s_{eff}[\psi]$, i.e. those which are relevant in the critical
 region.
   In the last step we find a Hamiltonian ${\cal
 H}_s[\psi]$, such that the Legeandre transform of the functional
\begin{eqnarray}
\label{Hs}
\log \int D\psi 
e^{\langle w|\psi\rangle}e^{-\beta {\cal H}_s[\psi]}
\end{eqnarray} 
generates the same vertex functions. The scaling behavior of the
vertex functions in the RPM is thus determined by the form of ${\cal
H}_s[\psi]$. In sec.5   we show that ${\cal H}_s[\psi]$ has the form
characterizing the Ising universality class.

At the end of this section let us summarize the foundations of the
effective theory.  The nonlocal parts of the connected number-density
correlation functions are given in Eqs.(\ref{etetcon})-
(\ref{etetcon1}). The generating functional for $\langle \psi({\bf
x}_1)... \psi({\bf x}_n)\rangle^{con}$ is defined in
Eq.(\ref{genfun2}), with the effective Hamiltonian ${\cal
H}_{eff}[\phi]$  given in Eq.(\ref{hef}).
\section{Origin of the  critical singularity}
 In the framework of the above coarse-grained approach the critical
 singularity of $ \tilde G_{\psi\psi} (0)$ was first found in
 Ref.\cite{ciach:00:0} in the WF approximation. The domain of validity
 of the equation for the spinodal line $ \tilde C_{\psi\psi} (0)=0 $
 was not determined, however. Before describing the origin of the
 critical singularity, we shall briefly summarize the properties of
 the charge-density correlation function, $G_{\phi\phi}(\Delta{\bf
 x})\equiv\langle\phi({\bf x})\phi({\bf x}+\Delta{\bf
 x})\rangle^{con}$, determined already in Ref.\cite{ciach:04:1}. The
 form of $G_{\phi\phi}$ is crucial for finding the domain of validity
 of the equation for the gas-liquid type spinodal line in the phase
 diagram. This is because the gas-liquid type phase separation can in
 principle be preempted by the charge ordering for $\tilde
 C_{\phi\phi}({\bf k}_b)\le 0$, where $\tilde C_{\phi\phi}({\bf k})=
 1/\tilde G_{\phi\phi}({\bf k})$, as is in fact the case on the
 simple-cubic lattice \cite{ciach:03:0,ciach:04:1}.
\subsection {$G_{\phi\phi}$ in the  self-consistent Hartree approximation}
\label{Hartree_self}

The field theory with
instability corresponding to $k_b\ne 0$ was first developed by
Brazovskii
\cite{brazovskii:75:0} for the $\varphi^4$ theory. The Brazovskii
theory leads to correct behavior of the correlation functions,
i.e. the artificial singularity found on the Gaussian level is
removed.  For the RPM we keep terms up to $O(\phi^6)$ in the
functional (\ref{hef}), because for low densities ${\cal A}_4<0$, and
for stability reasons the functional (\ref{hef}) can be truncated at
the term $\propto\phi^6$ in the theory with ${\cal A}_6>0$.
 In Ref.\cite{ciach:04:1} the Brazovskii
theory for $G_{\phi\phi}$ was considered for the RPM on the sc
lattice, where a shift of the line of continuous transitions to the
charge-ordered phase was found. Here we shall briefly describe the
continuum case, where the above transition disappears.

 In the one-loop Hartree approximation the correlation function is
 given by an infinite series of effectively one-loop diagrams, shown
 in Fig.4 (top). In Fourier representation a single loop in the
 second diagram in Fig.4 (top) represents the integral
\begin{equation}
\label{calG0}
{\cal G}_0=\int_{\bf
k} \tilde G^{0}_{\phi\phi}( k),
\end{equation}
where the integrand $\tilde G^0_{\phi\phi}( k)$ (see Eqs. (\ref{gpi})
and (\ref{calC})) can be written in the form
\begin{equation}
\label{gS}
\tilde G^0_{\phi\phi}( k)
=\frac{T^*}{\tau_0+ \Delta \tilde V(k)}.
\end{equation}
In the above $\Delta \tilde V( k)=\tilde V( k)-\tilde V(
k_b)$, where for $k=k_b$ $\tilde V( k)$ assumes a minimum, and
\begin{equation}
\label{tau0}
\tau_0=\frac{T^*\Big(1+\rho^*_0f_2(\rho^*_0)\Big)}{\rho^*_0}+\tilde V(k_b).
\end{equation}
  Other diagrams in Fig.4 (top)
 are products of ${\cal G}_0$. The self-consistent approximation is
 obtained, when in Eq.(\ref{calG0}) the integrand is replaced by the
 correlation function which is the result of the whole resummation
 (Fig.4, bottom). The resulting equation is then solved
 self-consistently.  In the above one-loop self-consistent Hartree
 approximation the ${\bf k}$-dependence of the correlation function is
 the same as given in Eq.(\ref{gS}), and only the critical parameter
 $\tau_0$ is rescaled.  We denote the rescaled critical parameter by
 $\tau$, and the correlation function in this approximation by $
 G^H_{\phi\phi}$. The corresponding ${\bf k}$-integral of $\tilde
 G^H_{\phi\phi}( k)$ is
\begin{equation}
\label{calGtau}
{\cal G}(\tau)\equiv \int_{\bf
k} \tilde G^H_{\phi\phi}( k)=\int_{\bf
k}\frac{T^*}{\tau+ \Delta \tilde V(k)}.\end{equation}
The self-consistent equation for $\tilde
G^H_{\phi\phi}( k)$ assumes the form (see Fig.4, 
bottom, and Ref.\cite{ciach:04:1})
\begin{figure}
\label{fig11}
\includegraphics[scale=0.5,angle=270]{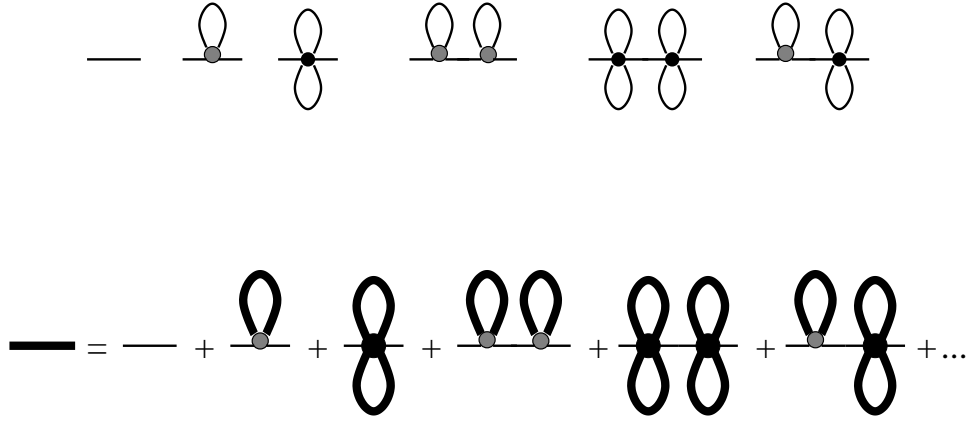}
\caption{Top: a few diagrams contributing to the charge-charge correlation 
function in the one-loop Hartree approximation. Bottom: Diagrammatic
representation of the self-consistent equation for the charge-charge
correlation function. The thick line represents the correlation
function $\tilde G^H_{\phi\phi}( k)$, thin line is the Gaussian
correlation function $\tilde G^0_{\phi\phi}( k)$, the shaded circle
 and the
bullet represent the
vertices $-{\cal A}_4$ and $-{\cal A}_6$ respectively, and the
 ${\bf k}$-integral is associated with each loop  (see also
Fig.2. }
\end{figure}
\begin{eqnarray}
\label{hartree}
\tilde G^H_{\phi\phi}( k)
=\Bigg[\tilde {\cal C}^{0}_{\phi\phi}(k)+ {\cal G}(\tau)
\Bigg(\frac{{\cal A}_4}{2}+
\frac{{\cal A}_6}{2^3}{\cal G}(\tau)\Bigg)\Bigg]^{-1}.
\nonumber
\end{eqnarray}
In the theory with the term $\propto \phi^8$ included in the effective
Hamiltonian, there is another contribution, proportional to ${\cal
A}_8{\cal G}(\tau)^3$, in the above equation.  From the definition
(\ref{tau0}) of $\tau_0$ we obtain the temperature corresponding to
the singularity of $\tilde G^H_{\phi\phi}(k_b)$ at $\tau=0$ (recall
that $\tilde V(k_b)<0$),
\begin{eqnarray}
\label{shiftT}
T^*=-\frac{\rho_0^*\tilde V(k_b)}{\Big[1+
{\cal G}(0)\Big(\frac{{\cal A}_4}{2}+
\frac{{\cal A}_6}{2^3}{\cal G}(0)\Big)\rho_0^*\Big]
\big(1+\rho^*_0f_2(\rho^*_0)\big)}.
\end{eqnarray}
 The above temperature is larger from zero for ${\cal G}(0)$ assuming
 a finite value, for example in the RPM on the sc lattice
 \cite{ciach:04:1}. For ${\cal
 G}(0)\to\infty$, however, $T^*\to 0$. 

In continuum there are two
 sources of divergency of ${\cal G}(\tau)$. The first, unphysical one
 is present for any $\tau$ and comes from the behavior of $\tilde
 V(k)$ for $k\to \infty$. The charge-density waves with $k\to \infty$
 would correspond to overlapping hard spheres, and should not be
 included. Following the standard procedure
\cite{brazovskii:75:0,fredrickson:87:0},
 we expand $\Delta \tilde V(k)$ in a Taylor series about $k=k_b$ and
 truncate the expansion in $\Delta k=k-k_b$.  The resulting
 approximate form agrees with $\tilde G^H_{\phi\phi}(k)$ for $k\approx
 k_b$, i.e.  for dominant fluctuations, and its behavior for $k\to
 \infty$ ensures finite values of ${\cal G}(\tau)$. In the following
 we shall consider the  integral (\ref{calGtau}) regularized as in Ref.
\cite{brazovskii:75:0}, 
\begin{eqnarray}
\label{calGr}
{\cal
 G}_r(\tau)=\int_{\bf
k}\frac{T^*}{\tau+ \frac{V''(k_b)}{2}\Delta k^2+O(\Delta k^3)}
=T^*k_b^2/(\pi\sqrt{2\tau\tilde V''(k_b)}) + ...,
\end{eqnarray}
where the subscript $r$ stands for 'regularized'. Note that for
reasonable regularizations (it is the neighborhood of $k_b$ that gives
the relevant contribution to the integral) the value of ${\cal
G}_r(\tau)$ depends on the regularization procedure only weakly (dots
in the above equation). 

The other, physical divergency of ${\cal
G}_r(\tau)$ occurs only for $ \tau=0$ and is associated with the
singular behavior of the integrand for ${\bf k}\to{\bf k}_b$,
i.e. with the dominant fluctuations $\tilde \phi({\bf k}_b)$
\cite{ciach:04:1}. However, the temperature corresponding to $\tau\to
0$ is $T^*\to 0$ (see Eqs.(\ref{shiftT}) and (\ref{calGr})), and for
$T^*>0$ we obtain in turn $\tau>0$, hence the charge-density
correlation function $\tilde G^H_{\phi\phi}(k)$ remains regular for
nonzero temperatures.  Instead, a first-order transition is found for
$\tau_0<0$
\cite{brazovskii:75:0}. In Ref.\cite{ciach:03:0} 
the fluctuation-induced first-order transition to the charge-ordered
phase was identified with formation of an ionic crystal. The fact that
the charge-density correlation function remains regular for nonzero
temperatures is crucial for the occurrence of the gas-liquid type
singularity in the continuum RPM.
\subsection{Critical singularity of $G_{\psi\psi}$}
In this section we describe the origin of the critical singularity of
$ \tilde G_{\psi\psi} (0)$, found already in
Ref. \cite{ciach:00:0}. Moreover, based on the properties of
$G_{\phi\phi}$ studied in Ref.\cite{ciach:04:1} and summarized above,
 we show that the equation for the
gas-liquid type spinodal line is well defined for any nonzero
temperature.

 Let us first consider the correlation functions 
$\langle\psi({\bf x}_1)...\psi({\bf x}_n)\rangle^{con}$.
 The higher-order contributions to the number-density
 correlation functions in Eq.(\ref{etetcon1}) will be considered
 later.   Contributions
 to the $n$-point correlation function for the field $\psi$ are given
 by diagrams contributing to the $2n$-point function for the field
 $\phi$, with $n$ pairs of external points identified, and multiplied
 by $a_1^{n}$, where $a_1$ is given in
 Eq.(\ref{a1}).
 In particular,  $G_{\psi\psi}$ (Eq.(\ref{rr3})) is
 given by connected diagrams of the four-point function for the field
 $\phi$, with the two pairs of external points identified with each
 other, and multiplied by $a_1^2$.
 The contribution to $\tilde G_{\psi
\psi}(0)$ leading to the critical singularity is given by
 an infinite series of diagrams which are of a form of chains of
 hyperloops connected by the vertices ${\cal A}_4$ \cite{ciach:00:0} 
(Fig.5). The sum of the corresponding geometric series has the form
\begin{equation}
\label{bareG}
\tilde G^b_{\psi\psi}(k)= 4\tilde g(k) 
\left[ 1+ {\cal A}_4\tilde g(k)\right]^{-1}a_1^2,
\end{equation}
 where $\tilde g(k)$ is a Fourier transform of the function
 representing the hyperloop, i.e. a sum of all connected diagrams of
 the four-point function with two pairs of external points identified,
 which cannot be split into two distinct diagrams by splitting a
 single vertex ${\cal A}_4$ into two two-point vertices, and the sum
 is divided by 2 (symmetry factor). Another words, $g(r)$, represented
 graphically by the hyperloop in Fig.5, has no contributions which are
 of a form of chains. We introduced a superscript $b$ to distinguish
 the function given in Eq.(\ref{bareG}) from the exact form of the
 correlation function defined in Eq.(\ref{rr3}).

Eq. (\ref{bareG}) stands a singular contribution to
 $\tilde G_{\psi\psi}$ for $k\to 0$ either when $\tilde g(0)\to
 \infty$ or when
\begin{equation}
\label{sin_gas-li}
1+{\cal A}_4 \tilde g(0)=0.
\end{equation}
  The above can be satisfied for ${\cal A} _4<0$.
Let us focus on the question for what temperatures $\tilde
g(0)$ is regular.
The lowest-order approximation for $g$ is represented by the first
loop in Fig.5 (bottom), and is given by
\begin{equation}
\tilde g_0(k)= {1\over 2}\int d {\bf r} G^0_{\phi \phi}(r)^{2} 
e^{i {\bf k r}}=  {1\over 2}\int d {{\bf k}'}
\tilde G^0_{\phi \phi}(k')\tilde G^0_{\phi \phi}(|{\bf k}-{\bf k}'|).
\end{equation}
$\tilde G^0_{\phi\phi}(k)$ is nonintegrable for $\tau_0\to 0$, because for $\tau_0=0$ 
$\tilde G^0_{\phi\phi}(k)$ diverges sufficiently fast when $k\to
k_b$. Thus, $\tilde G^{02}_{\phi\phi}(k)$ is nonintegrable as well,
and for $\tau_0\to 0$ we have $\tilde g_0(0)\to\infty$. This
singularity is associated with the charge-density waves with the
wavenumber $k_b$.
\begin{figure}
\includegraphics[scale=0.4]{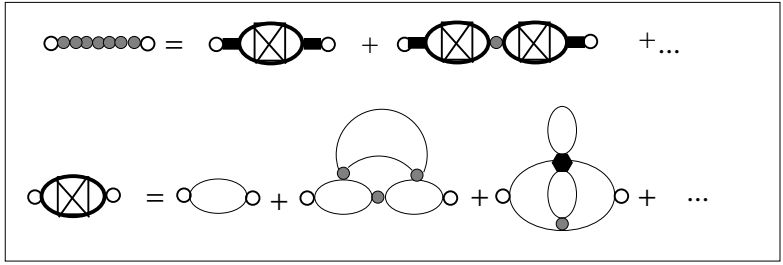}
\label{fig13}
\caption{Top: several diagrams contributing to the 
 correlation function $G^b_{\psi\psi}({\bf x}_1,{\bf
x}_2)$, with the external points ${\bf x}_1\ne {\bf
x}_2$ shown as open
circles. Black boxes represent $a_1$. In the approximation
corresponding to Eq.(\ref{bareG}), $G^b_{\psi\psi}({\bf x}_1,{\bf
x}_2)$ (represented by the pearl line) is given by an infinite series
of diagrams which have a form of chains of hyperloops.  Only diagrams
with the hyperloops connected by the vertices ${\cal A}_4$ (shaded
circle) are included in $G^b_{\psi\psi}$. Bottom: a few diagrams 
contributing to
the hyperloop $g({\bf x}_1,{\bf x}_2)$. The first contribution to
$g({\bf x}_1,{\bf x}_2) $ is $g_0({\bf x}_1,{\bf x}_2)=
G^0_{\phi\phi}({\bf x}_1,{\bf x}_2)^2/2$. }
\end{figure}
However, the divergency of $\tilde G_{\phi \phi}(k_b)$ is 
removed when the fluctuations are included within the Brazovskii
approach described in the previous subsection.
In a consistent  approximation the hyperloop
 in Fig.5, bottom, can be written in the form
\begin{equation}
\label{gfu}
g^H({\bf x}_1,{\bf x}_2)= \frac{1}{2}G_{\phi\phi}({\bf x}_1,{\bf
x}_2)^2+ ....,
\end{equation}
where $ G_{\phi\phi}({\bf x}_1,{\bf x}_2) $ is the charge-density
correlation function, and dots indicate all remaining contributions. Because  $
\tilde G_{\phi\phi}({\bf k})$ is regular, $\tilde g^H(0)$ is regular as well,
 when the
integrals are regularized for $k\to \infty$ in the way described in the
preceding subsection. Thus, the only singularity of $\tilde G_{\psi\psi}
(0)$ occurs at an infinite order in ${\cal A}_4$ and in this
approximation is given by Eq.(\ref{sin_gas-li}) with $g$ approximated
by $g^H$.The
function inverse to that given in Eq.(\ref{bareG})
 can be expanded about $k=0$, and we
obtain
\begin{equation}
\label{C00}
\tilde C^b (k)
=\frac{1}{\tilde G^{b}_{\psi\psi} (k)}= c_0 + c_2 k^2 + O(k^4),
\end{equation}
with $c_0$ and $c_2$ expressed in terms of $\tilde g^H(0)$ and $\tilde
g^{H''}(0)$ in a standard way.  Note that the form of $\tilde C^b (k)$ for
$k\to 0$ is the same as the form of the inverse Gaussian correlation
function in the Landau theory for simple fluids.  The above result
indicates that in the perturbation theory the separation into two
uniform ion-dilute and ion-dense phases is present at an infinite
order in ${\cal A}_4$ for $c_0=0$. The above discussion shows that  in the continuum
RPM
this separation is not preempted by the charge ordering.

We conclude this section by stressing that in the perturbation theory
we should make a resummation of infinite series of diagrams twice. The
first resummation, in the spirit of the Brazovskii
theory~\cite{brazovskii:75:0}, cures the artificial divergency of the
charge-density correlation function in the Gaussian approximation.
Critical singularity is not exhibited by any individual diagram, but
by an infinite series of diagrams having the form of chains of
hyperloops (Fig.5). The singularity is found only in the phase-space
region where ${\cal A}_4<0$.  Note that in the original functional
$\Delta\Omega^{MF}$ (Eq.(\ref{OmF})) the field $\tilde\phi({\bf k}_b)$
is critical, and the field $\eta$ is noncritical. Charge-density
fluctuations lead to essentially different roles of both fields for
${\cal A}_4<0$ -- the field $\phi$ is turned to be noncritical, and
the field $\eta$, or in fact $\psi\propto \phi^2$, becomes critical.
\section{Universality class of the RPM critical point} 
In this section we separate all diagrams contributing to the vertex
functions for the field $\psi$ into disjoint sets according to the
rules described below. The series of diagrams in each set is given by
an expression that can be represented by a secondary diagram of a
skeleton form. The secondary diagrams are again segregated into
disjoint sets. The series of secondary diagrams in each set can be
represented by a hyperdiagram. We show that the expressions
representing the hyperdiagrams are the same as the corresponding
expressions representing the diagrams in a certain model system
belonging to the Ising universality class. The one-to-one
correspondence between the hyperdiagrams representing the vertex
functions in the RPM, and the diagrams representing the vertex
functions in the model belonging to the Ising universality class
indicates that the singular part of the free-energy functional has the
Ising universality class form, up to additional terms associated with
corrections to scaling.
\subsection{Secondary diagrams}
 In the approximation studied in the previous section $g^H$ can be
 represented by a ``secondary diagram'' of the same topological form
 as the first diagram on the r.h.s. in Fig.5, bottom, but with the thin
 line representing $G^0_{\phi\phi}({\bf x}_1,{\bf x}_2)$ replaced by a
 'thick' line representing $G_{\phi\phi}({\bf x}_1,{\bf x}_2)
 $, as shown in Fig.6.
\begin{figure}
\includegraphics[scale=0.5]{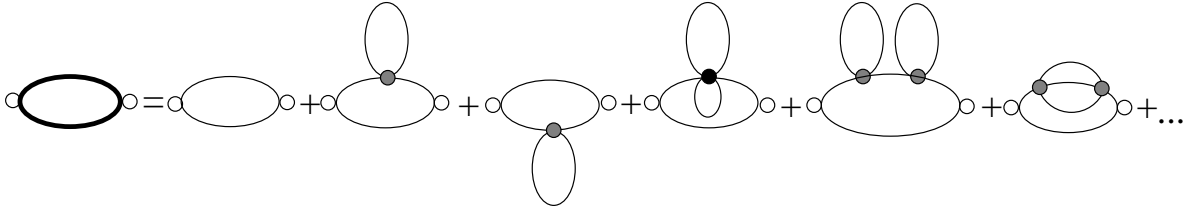}
\caption{One-loop secondary diagram contributing to 
$\langle\phi^2({\bf x}_1)\phi^2({\bf x}_2)\rangle^{con} $
 (l.h.s.), and the corresponding series of Feynman diagrams (r.h.s).
 Thick and 
 thin lines represent  $G_{\phi\phi}$ and  $G^0_{\phi\phi}$ respectively, and
 open, gray and black circles represent external points, ${\cal A}_4$ and 
${\cal A}_6$ respectively.
}
\end{figure}

 Similarly, all diagrams contributing to the connected correlation
 functions for the field $\psi({\bf x})=a_1\phi^2({\bf x})$, can be
 divided into disjoint sets. Each set consists of one skeleton
 diagram~\cite{zinn-justin:89:0} and of all diagrams obtained from
 this skeleton by insertions for any pair of points ${\bf x}',{\bf
 x}''$, connected by a thin line representing $G^0_{\phi\phi}({\bf
 x}',{\bf x}'')$, of subdiagrams which have a form of diagrams
 contributing to the two-point function $G_{\phi\phi}({\bf x}',{\bf
 x}'')$. For each set a series of all its diagrams can be represented
 by a secondary diagram.  The secondary diagram has the same
 topological form as the skeleton diagram, but all lines representing
 $ G^0_{\phi\phi}$ are replaced by the lines representing $
 G_{\phi\phi}$.  In a skeleton diagram there exists no pair of lines
 such that by cutting only these lines a subdiagram contributing to
 $G_{\phi\phi}$ could be extracted.  Individual secondary diagrams are
 regular, because $
\tilde G_{\phi\phi} ({\bf k})^n$  is regular and integrable when $
\tilde G_{\phi\phi} ({\bf k})$ is
 regularized for
 $k\to\infty$ as described above.  In the following we shall
always consider secondary diagrams. By doing so we automatically cure
the divergency associated with the unphysical singularity of $\tilde
G^0_{\phi\phi} ({\bf k}_b)$. 

 Because each vertex ${\cal A}_{2m}$ has an even number of
 $\phi$-legs, and in diagrams contributing to $\langle\psi({\bf
 x}_1)...\psi({\bf x}_2)\rangle^{con}$ an even number of lines
 representing $G_{\phi\phi}$ is connected with each external point,
 all diagrams contributing to $\langle\psi({\bf x}_1)...\psi({\bf
 x}_n)\rangle^{con}$ are 1PI. The secondary (skeleton) diagrams
 contributing to $\langle\psi({\bf x}_1)...\psi({\bf
 x}_2)\rangle^{con}$ consist of loops containing $n\ge 2$ thick lines,
 and the loops are connected by the vertices ${\cal A}_{2m}$ with
 $m\ge 2$. On the other hand, all diagrams contributing to $\langle
 \phi({\bf x}_1)... \phi({\bf x}_n)\rangle^{con}$ with ${\bf x}_1\ne
 ...\ne{\bf x}_n$ are one particle reducible (1PR), i.e. two separate
 diagrams can be obtained by cutting a single line. 

In the following  subsections we consider vertex functions and the
 free-energy
functional in the perturbation expansion in the vertices ${\cal
A}_{2m}$. We shall segregate the corresponding secondary diagrams into
 different disjoint
sets, associated with different contributions to the vertex functions.
  We focus on the general theory with ${\cal A}_4<0$ and ${\cal
A}_{2n}>0$ for $n\ge 3$. We first analyze the lowest-order
approximation.
\subsection{Perturbation expansion at the zeroth-order in the vertices
 ${\cal A}_{2m}$ with $m\ge 3$} 

Let us consider diagrams that contain
 loops connected only by the vertices ${\cal A}_4$. We shall first
 consider a general vertex part of the form of a loop with $n\ge 3$
 lines. We show that such vertex parts are negligible in the critical
 region, and that the free energy-functional (\ref{Gamma_exp3}) has
 the same form as in a model system described by a certain Hamiltonian
 of the Gaussian form.

Let us consider a general  secondary diagram contributing
to the $n$-th order vertex function. At the one-loop order we have
\begin{eqnarray}
\label{vertexvanish}
F^{one-loop}_{n}({\bf x}_1,...,{\bf x}_n)\propto  
 G_{\phi\phi}({\bf x}_1-{\bf x}_2)...
 G_{\phi\phi}({\bf x}_n-{\bf x}_1) ,
\end{eqnarray} 
and the corresponding loop is shown
  in Fig.7.
\begin{figure}
\label{fig14}
\includegraphics[scale=0.5]{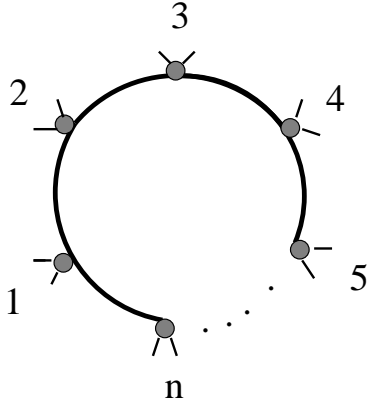}
\caption{A loop with $n$ lines  ($n\ge 3$) 
 contributing to the $n$-point vertex function for the field $\psi$,
 $F_{n}({\bf x}_1,...,{\bf x}_n)$. Lines represent
$G_{\phi\phi}$. For the above contribution we have
$\int_{{\bf x}_1}...\int_{{\bf x}_n}
F^{one-loop}_{n}({\bf x}_1,...,{\bf x}_n)
\propto \tilde G_{\phi\phi}(0)^n=0$.
}
\end{figure}
For $F_{n}({\bf x}_1,...,{\bf x}_n)$ integrated over all
 arguments we find
at the one-loop order
\begin{eqnarray}
\label{vertexvanishint}
\int_{{\bf x}_1}...\int_{{\bf x}_n}
F^{one-loop}_{n}({\bf x}_1,...,{\bf x}_n)\propto   \nonumber \\
\int_{{\bf x}_1-{\bf x}_2}...\int_{{\bf x}_n-{\bf x}_1}
 G_{\phi\phi}({\bf x}_1-{\bf x}_2)...
 G_{\phi\phi}({\bf x}_n-{\bf x}_1) \propto
\tilde G_{\phi\phi}(0)^n=0.
\end{eqnarray} 
 The last equality follows from the fact that $\tilde
 G_{\phi\phi}^0(0)=0$ (see Eqs.(\ref{phitad}) and (\ref{vfrpm})), and
 any diagram in the perturbation expansion of $ \tilde
 G_{\phi\phi}(0)$ is proportional to some power of $ \tilde G^0
 _{\phi\phi}(0)=0$. This is because all diagrams contributing to
 $\tilde G_{\phi\phi}$ are 1PR. Note that the above property indicates
 short-range of the charge-density correlations in the uniform phase.
 For a secondary diagram with any number of loops the
 corresponding contribution to $\int_{{\bf x}_1}...\int_{{\bf
 x}_n}F_{n}({\bf x}_1,...,{\bf x}_n)$ is also proportional to $\tilde
 G_{\phi\phi}(0)=0$ for $n\ge 3$.

 Let us consider $\Gamma_{eff}$, the 
 generating functional for
the vertex functions $F_{n}$ (see
Eq.(\ref{Gamma_exp3})). Because fluctuations relevant for the phase
separation vary very slowly in space, let us focus on $\psi({\bf
r})=\psi=const$, in which case we have
\begin{eqnarray}
\label{Gamma_exp00}
&\beta \Gamma_{eff}[\psi]=- \int_{\bf x}F_{1}({\bf
x}) \psi +\frac{1}{2}\int_{{\bf x}_1}\int_{{\bf x}_2}
C_{\psi\psi}(\Delta{\bf x})
\psi^2  \nonumber \\
&+ \sum_{n>2}
\int_{{\bf x}_1}...\int_{{\bf x}_n}
\frac{F_{n}({\bf x}_1,...,{\bf x}_n)}
{n!}\psi^n. 
\end{eqnarray} 
  Let us consider the last term in the above equation.  There are no
  zero-loop contributions to $F_{n}$ with $n\ge 3$ when the
  hypervertices ${\cal A}_{2m}$ with $m\ge 3$ are absent. As shown
  above, at a higher number of loops the last term in
  Eq.(\ref{Gamma_exp00}) vanishes, and only the first two terms
  remain.  From the resulting form of Eq.(\ref{Gamma_exp00}) it
  follows that neglecting the hypervertices ${\cal A}_{2n}$ with $n\ge
  3$ is analogous to  the Gaussian approximation,
  provided that the fluctuations with wave numbers other than $k= 0$
  are disregarded. The role of the $n$-point vertex functions
  $F_n({\bf k}_1,...,{\bf k}_n)$, which for $k_i\to 0$ behave as
  $\propto k_1^{2}...k_n^{2}$ will be discussed in section
  \ref{corr_sc}, where we briefly comment on all the irrelavant
  contributions to the vertex functions.

Let us determine the contribution to $G_{\psi\psi}$ coming from the
secondary diagrams containing loops with more than two lines.
 As an example consider the secondary diagram shown in
 Fig.8 (top).  In Fig.8 (bottom) the
 hyperdiagram, denoted by $f({\bf x}_1,{\bf x}_2)$, and obtained by a
 resummation of all diagrams of the above type is shown. We find
\begin{eqnarray}
\label{ver_van}
\int_{{\bf x}_1}\int_{{\bf x}_2}e^{i{\bf k}\cdot({\bf x}_1-{\bf x}_2)}
f({\bf x}_1,{\bf x}_2)=\tilde f(k)V= \\ \nonumber
\tilde G^{b}_{\psi\psi}(k)^2\tilde G_{\phi\phi}(k)
\tilde G_{\phi\phi}(0)^2\int_{{\bf x}'_2-{\bf x}'_3}
 G^{b}_{\psi\psi}({\bf x}'_2-{\bf x}'_3)
 G_{\phi\phi}({\bf x}'_2-{\bf x}'_3)=0,
\end{eqnarray}
where we used the fact that $\tilde G_{\phi\phi}(0)=0$.
From Eq.(\ref{ver_van}) we see that the
 above contribution to $\tilde G^{b}_{\psi\psi}(k)$ can be neglected.
 Similarly, all other diagrams of $\tilde G_{\psi\psi}(k)$, containing
 loops with more than two lines, are proportional to some power of
 $\tilde G_{\phi\phi}(0)=0$. Hence, when the hypervertices ${\cal
 A}_{2n}$ with $n>2$ are not included, the singular part of
 $G_{\psi\psi}$ is given by a sum of secondary diagrams of a form of a
 chain of loops with only two lines (Fig.9), and the inverse correlation
 function can be approximated by Eq.(\ref{C00}).
\begin{figure}
\label{Gro_van}
\includegraphics[scale=0.4,angle=270]{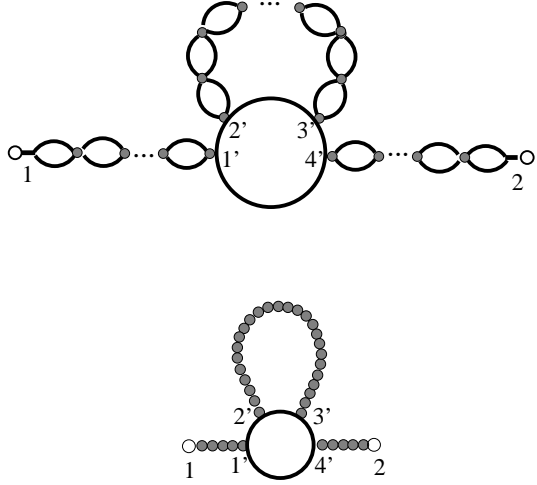}
\caption{Top: A secondary diagram contributing to $G_{\psi\psi}$ 
and containing 
the loop of the form (\ref{vertexvanish}).
 Lines represent $G_{\phi\phi}$, 
gray circles represent ${\cal A}_4$ and black boxes connecting the
external points with the loops represent $a_1$ (Eq.\ref{a1}).
 Bottom: A hyperdiagram
obtained after a resummation of all diagrams of the form shown above. 
The corresponding expression in Fourier representation is given in 
Eq.(\ref{ver_van})
}
\end{figure}

\begin{figure}
\label{Gro_van2}
\includegraphics[scale=0.4]{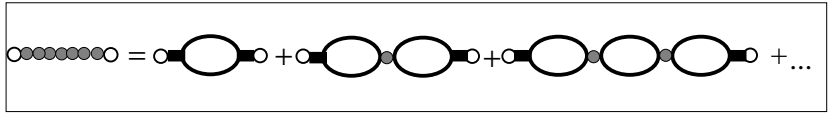}
\caption{Secondary diagrams contributing to $G_{\psi\psi}$
 (right-hand side)
 at the zeroth-order in the vertices ${\cal A}_{2n}$ with $n>2$.
 $G^b_{\psi\psi}$ is obtained after a resummation of all diagrams of
 the above form, and is represented by the pearl line. Symbols are
 explained in the caption of the previous figure. }
\end{figure}

 At the end let us consider the one-point function, i.e. the average
 density. The  secondary diagrams contributing to
 $\langle\psi({\bf x})\rangle =a_1\langle \phi^2({\bf x})\rangle$ 
at the zeroth order in
 the hypervertices ${\cal A}_6$ and ${\cal A}_8$ are shown in
 Fig.10,  and the corresponding 
 density shift is given by 
\begin{equation}
\label{psi_shi}
\langle \psi({\bf x})\rangle=
a_1{\cal G}_r(\tau)\Big[1-{\cal A}_4 \tilde g(0)
+(-{\cal A}_4 \tilde g(0))^2 +... (-{\cal A}_4\tilde
g(0))^n\Big]=
\tilde G^b_{\psi\psi}(0) F^0_1
,
\end{equation} 
where Eq.(\ref{calGtau}) with $G^H_{\phi\phi}$ replaced by $G_{\phi\phi}$,
 and  Eq.(\ref{bareG}) have been used, and we have introduced
\begin{equation}
\label{gamma010}
 F^0_1({\bf x})= F^0_1=\frac{ {\cal G}_r(\tau)}{4\tilde g(0)a_1}.
\end{equation}
Note that $ F^0_1({\bf x})$  plays a role analogous to the external field.
\begin{figure}
\label{fig15}
\includegraphics[scale=0.4]{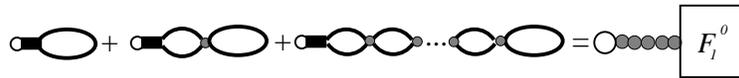}
\caption{ Diagrams contributing to the average density 
at the point 
shown as an open circle, and the hyperdiagram representing the sum
of these  diagrams. The black box represents $a_1$ (Eq.\ref{a1}).
 The pearl line represents $G^b_{\psi\psi}$ (Eq.(\ref{bareG}))
 and $F^0_1 $
inside the box is given in Eq.(\ref{gamma010}).}
\end{figure}
The above results lead  to 
\begin{equation}
\label{Gamma001}
\beta \Gamma_{eff}[\psi]=-\langle F_1^0|\psi\rangle+
\frac{ 1}{2}\langle\psi| C^b_{\psi\psi}|\psi\rangle, 
\end{equation}
where $  F^0_1$ and $\tilde C^b_{\psi\psi}(k)$ are given in
 (\ref{gamma010}) and
(\ref{C00}) respectively.  The same form of the free-energy functional
is obtained  in simple fluids with the coarse-grained Hamiltonian 
\begin{equation}
\label{calH}
\beta{\cal H}[\psi] =
\frac{ 1}{2}\int_{\bf
k}\tilde C^b_{\psi\psi}(k)\tilde\psi({\bf k})\tilde\psi(-{\bf k})-
\int_{\bf x} F^0_1\psi({\bf x})+\beta{\cal H}_{int},
\end{equation}
with the term ${\cal H}_{int}$ neglected. 
\subsection{Even and odd secondary diagrams}
The above observations indicate that 
the secondary diagrams contributing to the vertex functions for the
field $\psi$ can be divided into two disjoint sets. In the first set,
which we call the set of even diagrams, any contribution to 
the vertex function
$F_{n}({\bf x}_1,...,{\bf x}_n)$ is
represented by secondary diagrams such that any pair of points is
either connected by an even number of lines, or is not directly
connected. Such diagrams consist of chains of two-line loops, as in
Fig.9 and in Figs.11-14 below. The diagrams in the other set contain loops
such that there exist pairs of points connected by an odd number of
lines. In particular, loops of the type shown in Fig.7, where points
${\bf x}'$ and ${\bf x}''$ are connected by a single line representing
$G_{\phi\phi}({\bf x}'-{\bf x}'')$, belong to this set of diagrams. We
have shown that for $\psi$ independent of the space position such
diagrams give a vanishing contribution to the free energy. In sec.5E  we
shall show that the odd diagrams in which some pairs of points 
are connected by three or more $\phi$-lines lead to a modification of the
coupling constants in the corresponding coarse-grained Hamiltonian 
(\ref{calH}) and to corrections to scaling, but do not change
the universality class, which is determined by the even
diagrams. 

Let us return to the function $g$ representing the hyperloop (see
Fig.5).  Note that except from the secondary diagram of the same
topological form as the first diagram shown in Fig.5, bottom, all
remaining secondary diagrams contributing to $g$ contain pairs of
points connected by an odd number of lines and belong to the set of
the odd diagrams. Hence, the approximation for $g$ given in
Eq.(\ref{gfu}) is justified as long as we limit ourseles to the set of
even diagrams.

The functional (\ref{Gamma001}) was obtained by a resummation of all
secondary diagrams with the two-line loops connected only by the
vertices ${\cal A}_4$. The question which arises at this point is the
role of the vertices ${\cal A}_6$ and ${\cal A}_8$.  For clarity of
presentation we shall focus in the following subsection only on the
even secondary diagrams. We shall determine the scaling behavior of
the corresponding contribution to the vertex functions.
\subsection{Even diagrams containing vertices  ${\cal A}_6$ and ${\cal A}_8$}
Let us first consider the three- and four-point vertex functions at
the first order in the vertices ${\cal A}_6$ and ${\cal A}_8$, and at
an arbitrary order in ${\cal A}_4$. If we include only even diagrams,
then the associated connected correlation functions are shown in
Fig.11 (left). By summing the infinite series of all diagrams of the
above form we obtain the hyperdiagrams shown in Fig.11 (right).
\begin{figure}
\label{three_four}
\includegraphics[scale=0.4,angle=270]{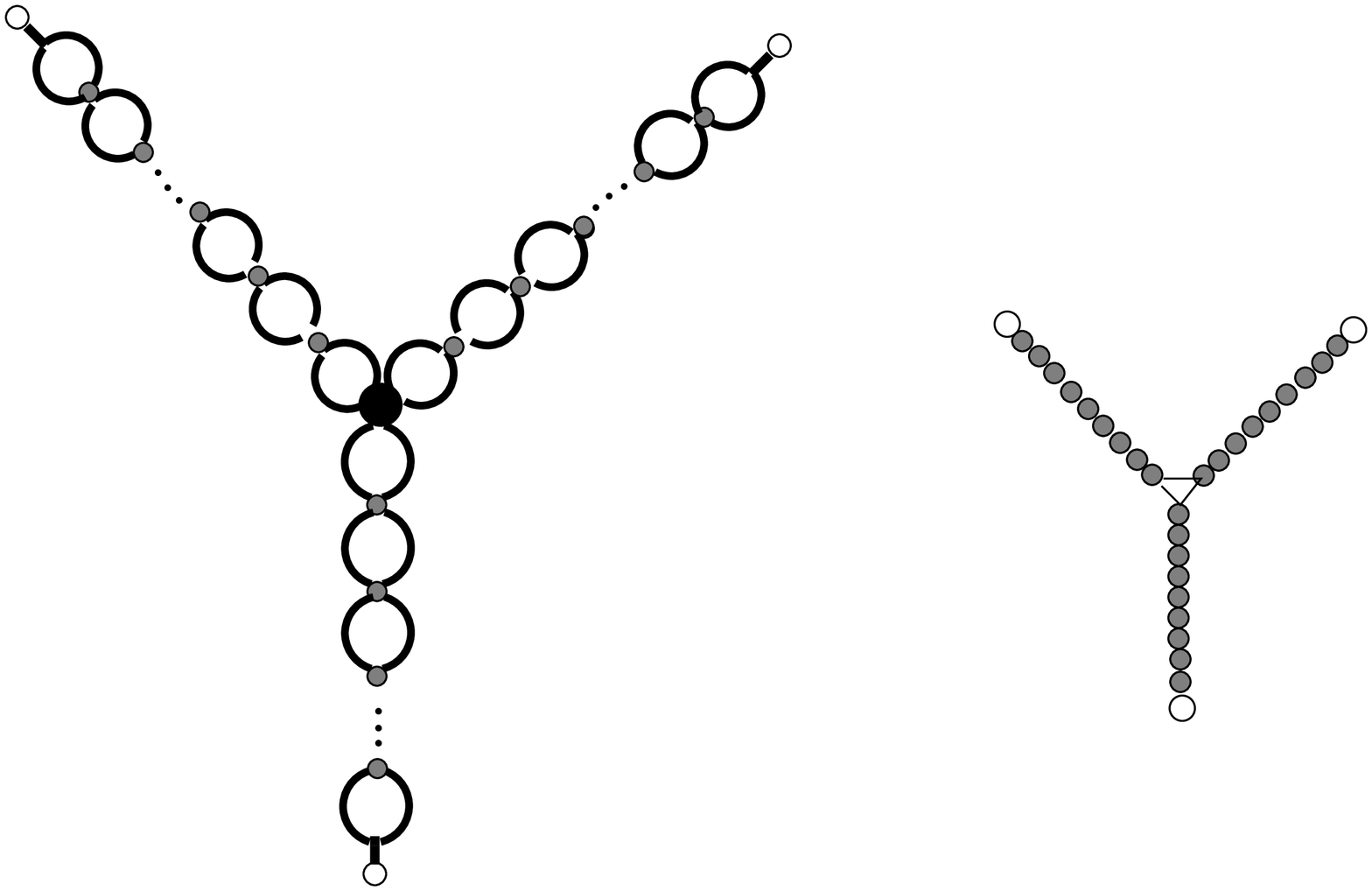}\\
\includegraphics[scale=0.4,angle=270]{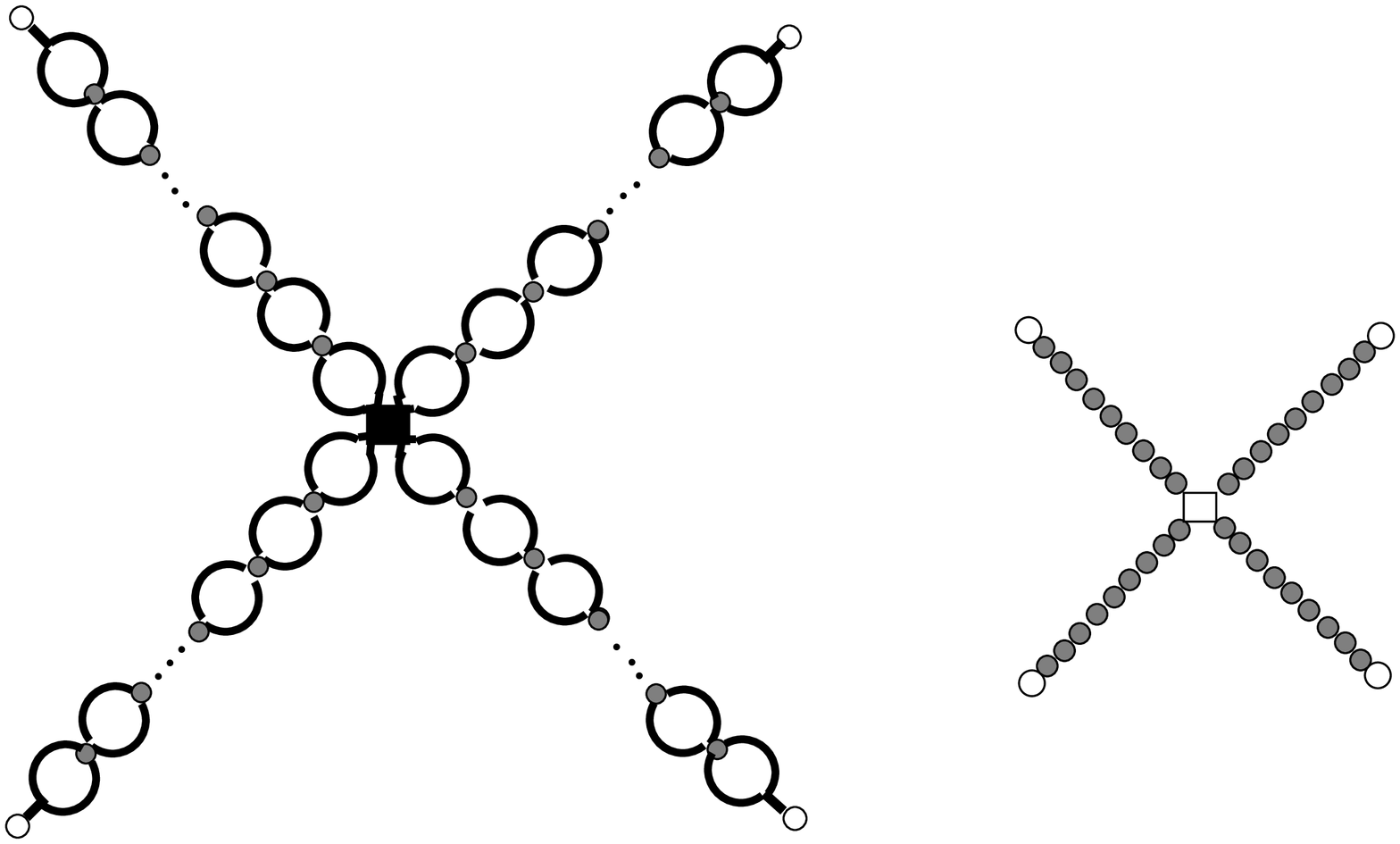}
\caption{Top: a secondary diagram contributing to
 the connected three-point 
correlation function at the first order in ${\cal A}_6$ (left) and the
hyperdiagram resulting from the resummation of all diagrams of the
above form.  Bottom: a diagram contributing to the connected
four-point correlation function at the first order in ${\cal A}_8$
(left) and the hyperdiagram resulting from the resummation of all
diagrams of the above form. The black square represents ${\cal A}_8$,
the triangle represents the three-point vertex $F^0_{3}$ and the open
square represents the four-point vertex $F^0_{4}$. The other symbols
are the same as in the previous figures.}
\end{figure}
Any other secondary (i.e. of a skeleton form) diagram at the first order
 in ${\cal A}_{6}$ or ${\cal A}_{8}$, contributing to
the respective  correlation functions, belongs to the set of odd
diagrams.  The three- and four- point vertex functions at the first
order in ${\cal A}_6$ and ${\cal A}_8$ respectively are just given by
\begin{equation}
\label{gamma003}
\tilde F^0_3({\bf k}_1,{\bf k}_2,{\bf k}_3)
=F^0_3=\frac{{\cal A}_6}{a_1^{3}}
\end{equation}
and
\begin{equation}
\label{gamma004}
\tilde  F^0_4({\bf k}_1,{\bf k}_2,{\bf k}_3,{\bf k}_4)=F^0_4
= \frac{{\cal A}_8}{a_1^{4}}.
\end{equation}
Due to the translational invariance in the real space, the Fourier
transforms are defined as
\begin{equation}
\label{mulFou}
(2\pi)^d\delta\big(\sum_i^n{\bf k}_i)\tilde  F_n({\bf k}_1,...,{\bf k}_n)=
\int_{{\bf x}_1}...\int_{{\bf x}_n}F_n({\bf x}_1,...,{\bf x}_n)
\exp\Big(i\sum_j^n{\bf x}_j{\bf k}_j\Big).
\end{equation}
 The hyperdiagrams on the right in Fig.11 are
of the same form as the corresponding diagrams contributing to the
connected correlation functions generated by the functional (\ref{Hs})
with the Hamiltonian (\ref{calH}), where
\begin{equation}
\label{Hin}
\beta{\cal H}_{int}=\int_{\bf r} \Bigg(\frac{F^0_3}{3!} 
 \psi^3({\bf r})+ \frac{F^0_4}{4!} \psi^4({\bf r})\Bigg),
\end{equation}
when calculated  at the first-order in
 the couplings $F^0_3$ and $ F^0_4$. 

At the first order in ${\cal A}_8$ there exists a contribution to the
connected three-point correlation function of the same form as shown
in Fig.11, top, left, but with the vertex ${\cal A}_6$ replaced by
${\cal A}_8{\cal G}_r/2$. However, the corresponding diagram belongs
to the set of diagrams shown in Fig.12, left, where the number of
two-line loops representing $g$ is arbitrary, from zero to infinity.
Similarly, the contribution to the two-point connected correlation
function with one vertex ${\cal A}_4$ replaced by ${\cal A}_6{\cal
G}_r/2$ belongs to the set of diagrams shown in Fig.13, left, and the
corresponding contribution with one vertex ${\cal A}_4$ replaced by
${\cal A}_8{\cal G}^2_r/2^3$ belongs to the set of diagrams shown in
Fig.14, left, bottom.
\begin{figure}
\includegraphics[scale=0.4]{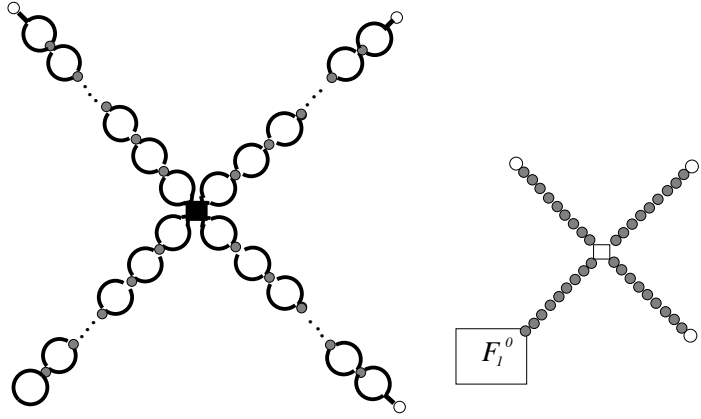}
\caption{Left: a secondary diagram contributing to the connected three-point 
correlation function at the first order in ${\cal A}_8$ 
 and the hyperdiagram resulting from the
resummation of all diagrams of the above form.  The symbols are the
same as in the previous figures.}
\end{figure}

 Let us consider an arbitrary vertex function for the field $\psi$,
 and focus on a particular contribution of a form of an even 1PI
 secondary diagram containing a given number of the vertices ${\cal
 A}_6$ and ${\cal A}_8$. Such a diagram consists of these vertices
 connected by chains of two-line loops, and the loops in the chain are
 connected by ${\cal A}_4$. Some of the chains of loops may end with
 the loop representing ${\cal G}$, as in Fig.12, left. By changing the
 number of loops in the chains we obtain other diagrams. All diagrams
 obtained in this way form a set. By a resummation of all diagrams
 belonging to this set we obtain a contribution to the considered
 vertex function, which can be represented by a hyperdiagram. The
 topological form of the hyperdiagram is given by any member of the
 set, if the chains of loops connecting the vertices ${\cal A}_6$
 and/or ${\cal A}_8$ are replaced by the pearl lines, as shown in
 Figs.11-15.

It is instructive to consider a few examples.  The even secondary
diagrams contributing to $\tilde G_{\psi\psi}$ and containing a single
vertex ${\cal A}_6$ and ${\cal A}_8$, are shown in Figs.13
and 14
 (left), respectively. By summing the infinite series of
all diagrams of the above form we obtain the hyperdiagrams shown in
Figs.13 and 14 (right), with the pearl lines
representing the 'bare' correlation function (\ref{bareG}). The
diagrams contributing to the connected four-point function at the
second order in ${\cal A}_8$ are shown in Fig.15.

 All even secondary diagrams can be separated into disjoint sets
 obtained in the way described above.  By a resummation of all
 diagrams belonging to any given set we obtain a hyperdiagram.  For
 any hyperdiagram contributing to the vertex function in the RPM there
 exists a diagram of the same topological structure and representing
 the same expression, contributing to the vertex function in the model
 given by the Hamiltonian (\ref{calH}) and (\ref{Hin}), and vice
 versa. This is just a topological property related to the possibility
 of connecting three- and four point vertices by lines, when the
 functional forms of the pearl line in the hyperdiagrams and the line
 in the diagrams corresponding to the above Hamiltonian are the same.
 We have thus reduced the question of the critical behavior of the
 vertex functions determined by the even diagrams to the question
 of the critical behavior of the system described by the Hamiltonian
 (\ref{calH}) with (\ref{Hin}),
\begin{figure}
\label{fig16}
\includegraphics[scale=0.5]{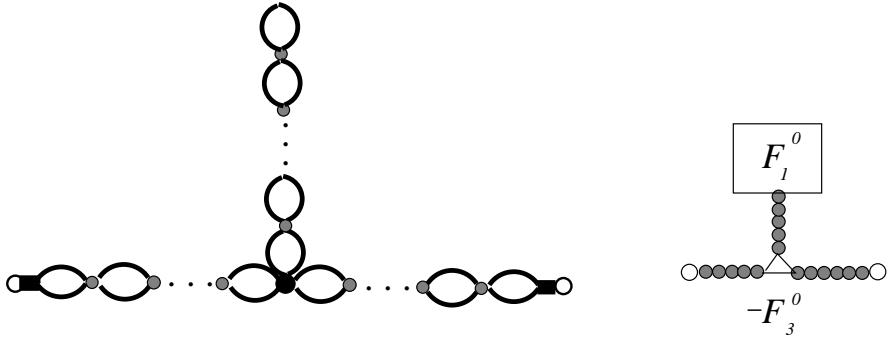}
\caption{A diagram contributing to the number-density correlation
 function at the first order in ${\cal A}_6$ represented by the bullet
 (left). The contribution to $G_{\psi\psi}$, given by the infinite
 series of such diagrams with all numbers of hyperloops, is shown
 schematically (right) as a hyperdiagram, where the pearl line
 represents the bare function (\ref{bareG})  and the open triangle
 represents $-F^0_3$(see Fig\ref{fig13}).
}
\end{figure}
\begin{figure}
\label{fig16a}
\includegraphics[scale=0.5]{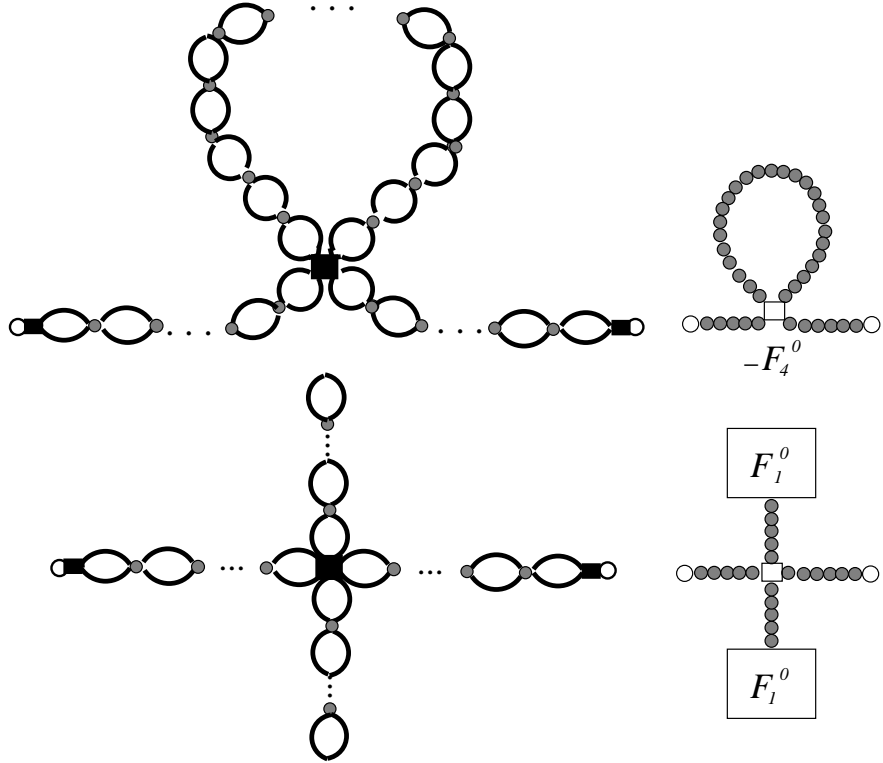}
\caption{Left: secondary diagrams contributing to the number-density 
correlation function at the first order in ${\cal A}_8$, represented
by the black square. The contributions to $G_{\psi\psi}$, given by the
infinite series of such diagrams with all numbers of loops
representing $g^H$, are shown schematically (right) as hyperdiagrams,
where the pearl line represents the bare function (\ref{bareG}) and
the open square represents $- F^0_4$).  }
\end{figure}
\begin{figure}
\label{fig16b}
\includegraphics[scale=0.5,angle=270]{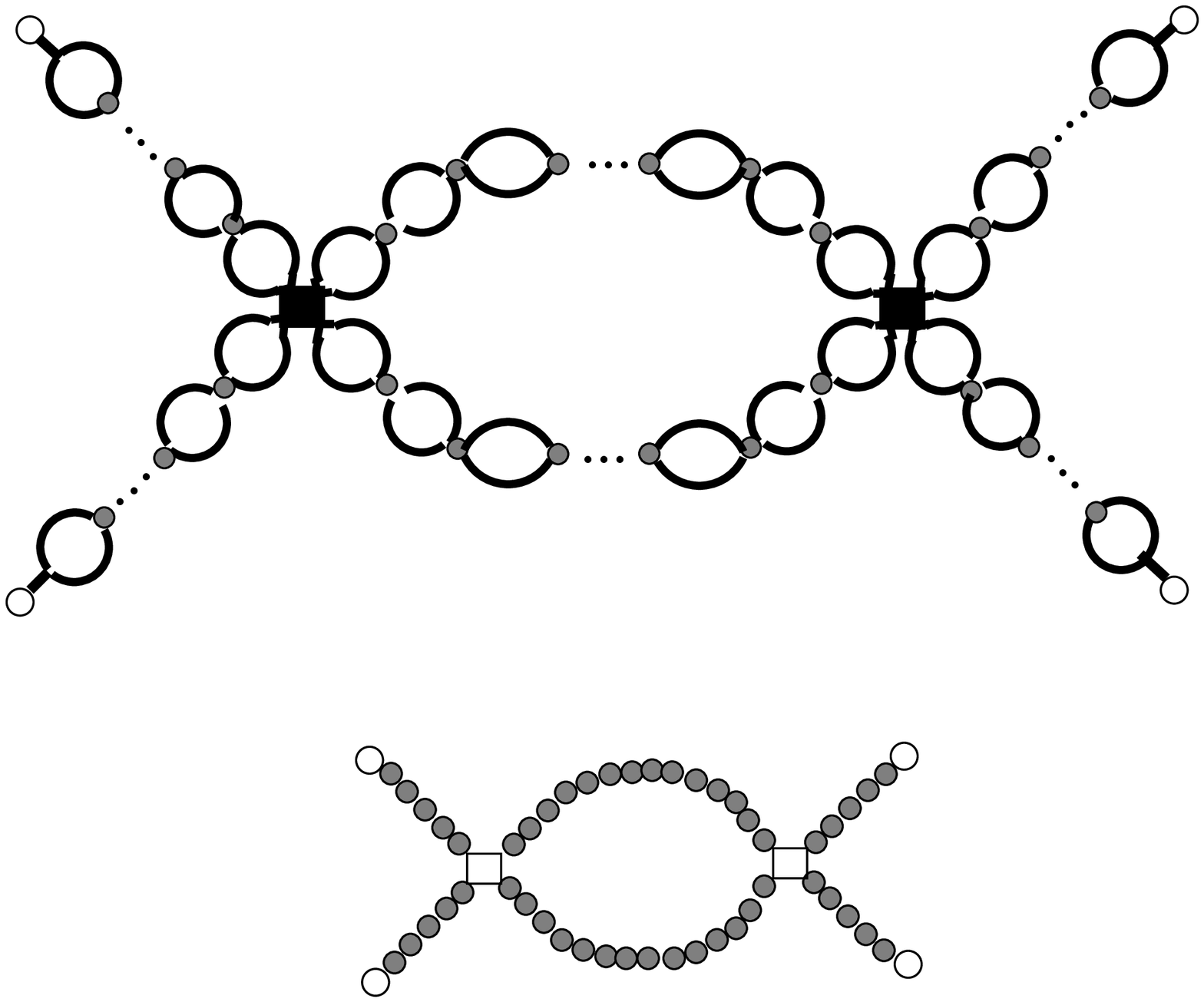}
\caption{Top: A secondary diagram contributing to the connected 
four-point number-density correlation
 function at the second order in ${\cal A}_8$ (black square).  Bottom:
 hyperdiagram representing an infinite series of all such diagrams. The
 pearl line represents the bare function (\ref{bareG}).}
\end{figure}
where $\tilde C^b_{\psi\psi}(k)$, $F^0_{1}, F^0_3 $ and $F^0_4$ are
given in Eqs. (\ref{C00}), (\ref{gamma010}), (\ref{gamma003}) and
(\ref{gamma004}) respectively.  We could include also higher-order
vertices ${\cal A}_{2m}$ with $m>4$, and we would obtain in the same way
additional terms $\propto\psi^m$ in Eq.(\ref{Hin}). However, such
terms are irrelevant in the RG sense
\cite{amit:84:0,zinn-justin:89:0}.

  By shifting the field,
\begin{eqnarray}
\psi=\varphi-\frac{F^0_{3}}{F^0_{4}},
\end{eqnarray}
we
can remove the term $\propto
\psi^3$ in Eq.(\ref{Hin}), and we obtain the standard 
form of the coarse-grained Hamiltonian
 representing the Ising universality-class, 
${\cal H}[\psi]/c_2={\cal H}^I[\varphi]+const.$, with
 ${\cal H}^I[\varphi]$
 given in Eq.(\ref{phi44}),
 where  the two relevant scaling fields are
\begin{eqnarray}
\label{H}
H_0=\Bigg(F^0_{1}+\frac{F^0_{3}}
{F^0_{4}}c_0-\frac{F^{03}_{3}}{3F^{02}_{4}}\Bigg)c_2^{-1}
 \end{eqnarray}
and
\begin{eqnarray}
t_0=\Bigg(c_0-\frac{F^{02}_{3}}{2F^{0}_{4}}\Bigg)c_2^{-1},
\end{eqnarray}
 and $u_0=F^0_{4}c_2^{-1}$.  Note that the scaling fields $H_0$ and
 $t_0$ are related to the temperature and the chemical potential
 (through $\rho_0^*$) in a somewhat complex way. Moreover, the values
 of the above parameters depend on ${\cal G}_r$ and $\tilde g(k)$,
 which in turn depend on the regularization procedure of the ${\bf
 k}$-integrals of $\tilde G_{\phi\phi}$ and its powers. We shall not
 study the nonuniversal properties in this work.

 The contributions of all even diagrams to the vertex functions for
 the field $\psi$ are of the same form as the vertex functions
 determined by the Hamiltonian given in Eq.(\ref{phi44}) and belonging
 to the Ising universality class. Hence, the scaling properties are
 also the same. We need to find out whether the odd-diagrams
 contributions to the vertex functions can alter the scaling
 properties.
\subsection{Odd diagrams containing vertices ${\cal A}_6$ and ${\cal A}_8$}
\label{corr_sc}
In this section we consider the secondary diagrams contributing to the
vertex functions for the field $\psi\propto \phi^2$, in which there
exist pairs of points connected by an odd number of lines representing
$G_{\phi\phi}$. We focus on the theory with the vertices ${\cal
A}_4<0$ and ${\cal A}_{2n}>0$ for $n\ge 3$. Our purpose is to show
that the odd diagrams give the contributions to the vertex functions
which either scale in the same way as the even diagrams, or are
associated with corrections to scaling. Because an even number of the
$\phi$-lines emanates from each vertex and each external point, and an
even number of lines is amputated from the external points in diagrams
contributing to the vertex functions for $\psi$, the vertices
connected by an odd number of lines form closed loops.
\subsubsection{Irrelavant odd diagrams}
  In sec.5.B we have already considered loops formed by single lines,
  as shown in Fig.7, where $\tilde F_n({\bf k}_1,...{\bf k}_n)\propto
  \prod_{i}^n\tilde G_{\phi\phi}(k_i)\propto
  k_1^{2}... k_i^{2}...k_n^{2}$ for $k_i\to 0$. Let us focus on a
  contribution to $\tilde F_n({\bf k}_1,...{\bf k}_n)$ that is of the
  form $\tilde F^{odd}_{\{j_i\}}({\bf k}_1,...{\bf k}_n)\propto
  k_1^{2j_1}... k_i^{2j_i}...k_n^{2j_n}$, where $j_i=0,1$ and
  $\sum_{i}j_i>0$, and in the corresponding secondary diagrams no pair
  of points is connected only by a chain of the two-line $g$-loops,
  and no loops representing ${\cal G}$ are present.  A series of such
  secondary diagrams for a given $\{j_i\}$ can be represented by a
  hyperdiagram containing no pearl lines. We have $\int_{{\bf x}_1}
  ...\int_{{\bf x}_n}F_{\{j_i\}}^{odd}({\bf x}_1,...{\bf x}_n)=0$,
  because $\sum_{i}j_i>0$.

By following the considerations
  described in the preceding subsection, we can segregate all
  secondary diagrams contributing to a particular vertex function into
  disjoint sets. In the secondary diagrams belonging to each set the
  vertices ${\cal A}_{6}$, ${\cal A}_8$ and $F_{\{j_i\}}^{odd}$ are
  connected by the chains of the two-line $g$-loops. All diagrams in any
  given set are of the same topological structure, when the chains of
 the  two-line loops are replaced by the pearl lines. The series of all
  diagrams in the given set is represented by a hyperdiagram. There is
 a one-to-one correspondence between the hyperdiagrams described above,
  and diagrams obtained in the perturbation expansion in the theory
  with the coarse-grained Hamiltonian (\ref{calH}) and with ${\cal
  H}_{int}$ (\ref{Hin}) supplemented with contributions of the form
  $O(\psi)\propto\int_{{\bf k}_1}...\int_{{\bf
  k}_n}\delta(\sum_i^n{\bf k}_i)
\prod_{i}^n\tilde\psi({\bf k}_i)k_i^{2j_i}$ with $n\ge 3$. 
 Note, however that in
the critical region the terms $O(\psi)$ correspond to irrelevant
operators in the RG sense
\cite{zinn-justin:89:0}.
\subsubsection{Relevant odd secondary diagrams}
Let us consider odd secondary diagrams contributing to the vertex
functions for the field $\psi$, other than those proportional to
$\tilde G_{\phi\phi}(0)=0$, which are irrelevant. In such diagrams
there exist closed loops in which neighboring pairs of points, ${\bf
x}_1$ and ${\bf x}_2$, are connected by subdiagrams contributing to
$G_3({\bf x}_1-{\bf x}_2)=\langle\phi^3({\bf x}_1)\phi^3({\bf
x}_2)\rangle$. The diagrams contributing to $G_3({\bf x}_1-{\bf x}_2)$
are shown in Fig.16.
\begin{figure}
\includegraphics[scale=0.35]{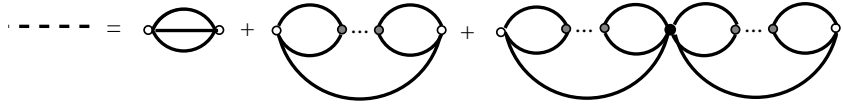}
\caption{Relevant odd secondary diagrams contributing to the correlation
 function $G_3({\bf
x}_1,{\bf x}_2)=\langle\phi^3({\bf x}_1)\phi^3({\bf x}_2)\rangle$.
 The dashed line represents the series of all
diagrams of the form shown on the r.h.s.}
\end{figure}
 Because $\tilde G_{\phi\phi}(k)\sim k^2$ for $k\to 0$, and the
 integrals are regularized for $k\to\infty$ as described in sec.4, the
 individual diagrams contributing to $\tilde G_3(0)$ are finite. The
 series of diagrams of the form shown in the second diagram on the
 r.h.s. in Fig.15  is also regular for the same reason.  In the theory
 with ${\cal A}_6>0$ the series of all diagrams shown on the r.h.s. in
 Fig.15 is regular for $k\to 0$, and we find $\tilde G_3(k)=
 g_3^0+g_3^2k^2+...$. Note the significant difference between the
 pearl line (Fig.9) obtained after the resummation of the even
 diagrams, and the dashed line obtained after the resummation of the odd
 diagrams (Fig.16). Because ${\cal A}_4<0$ and ${\cal A}_6>0$,
 only the pearl
 line can be singular for $k\to 0$.
\begin{figure}
\includegraphics[scale=0.5]{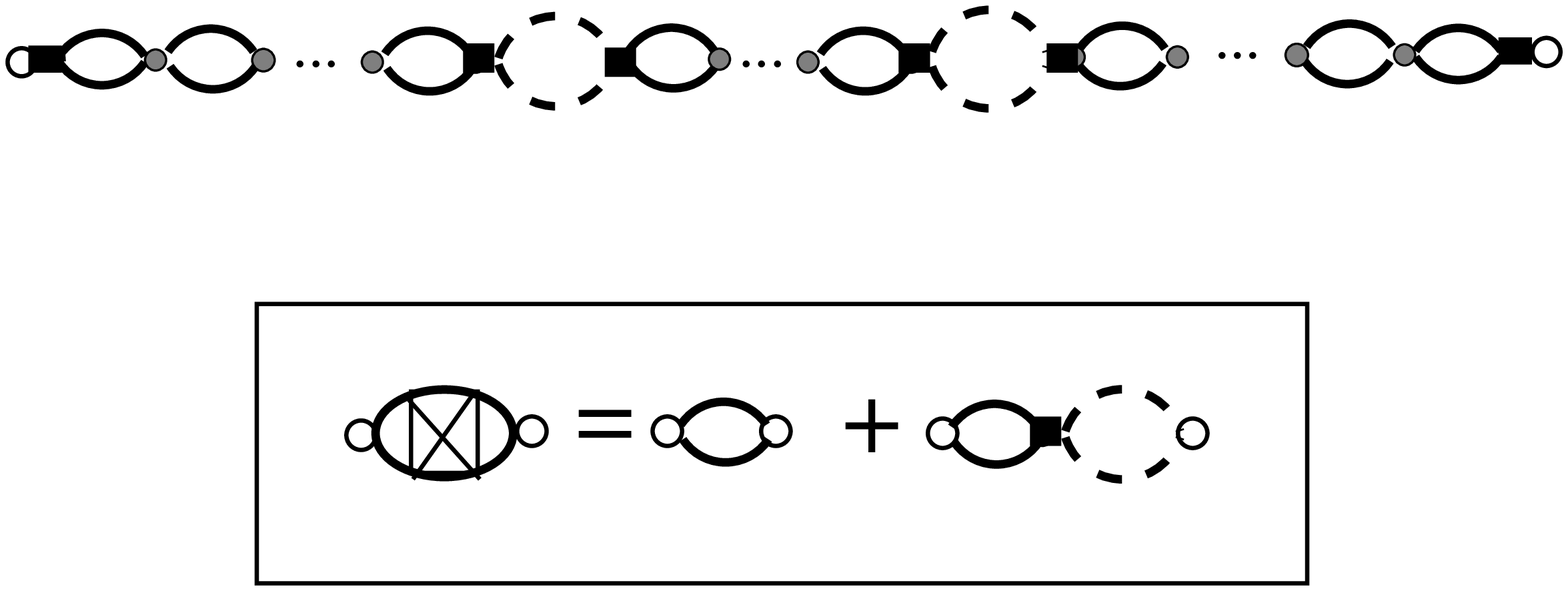}
\caption{Top: Relevant odd secondary diagram contributing to 
the two-point 
 correlation function for the field $\psi$.  The dashed lines
 represent the sum over all subdiagrams contributing to $G_3$, and the
 solid lines represent $G_{\phi\phi}$. The gray circles, black
 squares, open circles and black boxes represent ${\cal A}_4$, ${\cal
 A}_8$, the external points, and $a_1$ respectively. Bottom: secondary
 diagrams contributing to the hyperloop representing $g$, when the
 even as well as the relevant odd diagrams are included, and the
 vertices ${\cal A}_{2m}$ with $m>4$ are neglected.}
\end{figure}
In Fig.17, top, we show a relevant odd secondary diagram contributing
to the two-point correlation function for the field $\psi$. This
diagram represents in fact the series of all secondary diagrams that
contain subdiagrams contributing to $G_3({\bf x}-{\bf x}')$ for each
pair of the vertices ${\bf x}$ and ${\bf x}'$ connected by the dashed
line in Fig.17. We shall keep the name 'a secondary diagram' for such
a series of secondary diagrams.  The corresponding secondary diagrams
contributing to the hyperloop $g$ (see sec.IVB, Fig.5) are shown in
Fig.17, bottom, where the irrelevant diagrams are not included. In
this approximation we obtain for $\tilde G^b_{\psi\psi}(k)$ the same
expression as given in Eq.(\ref{bareG}), but with $\tilde g(k)$ given
by
\begin{equation}
\tilde g(k)=\tilde g^H(k)\Big(1+\frac{{\cal A}_8}{2}
\int_{\bf x}G_3({\bf x})^2e^{i{\bf k}{\bf x}}\Big),
\end{equation}
 where $g^H$ is given in Eq.(\ref{gfu}), and $G_3$ is shown in
 Fig.16. For $k\to 0$ we obtain the bare two-point correlation
 function $\tilde G^b_{\psi\psi}(k)$ of the same form as given in
 Eq.(\ref{C00}), but with modified coefficients $c_0$ and $c_2$.  Note
 that the secondary diagrams contributing to the bare function $\tilde
 G^b_{\psi\psi}(k)$ contain vertices other than ${\cal A}_4$, unlike
 in the case of the even diagrams. Note also that the secondary diagrams which
 contain chains of loops such that the two ends of the chain emerge
 from the same vertex, or one end of the chain represents ${\cal
 G}(\tau)$ (as is the case in Fig.14), are not included in the bare
 function $ G^b_{\psi\psi}$. As in the case of the even secondary
 diagrams, only linear chains, with no chain-like branches, contribute
 to $ G^b_{\psi\psi}$.
\begin{figure}
\includegraphics[scale=0.45]{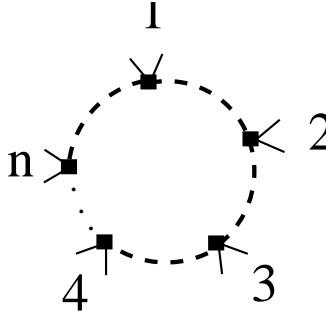}
\caption{Relevant odd  secondary diagram contributing to the $n$-point 
 vertex function for the field $\psi$. The black squares
 represent ${\cal A}_8$. The dashed line represents
 the three-point correlation function
 $G_3$. Thin lines represent the (amputated) $\phi$-lines.}
\end{figure}

 In Fig.18 we show the contribution to the $n$-point vertex function
 for the field $\psi$, $\tilde F^{odd}_n({\bf k}_1,...,{\bf k}_n)$, of
 the form $ \prod_i\tilde G_3(k_i)$. Note that $\tilde
 G_3(k)=g^0_3+g^2_3k^2+O(k^4)$ is finite for $k\to 0$, therefore the
 above contribution to $\tilde F^{odd}_n({\bf 0},...,{\bf 0})$, being
 of the form $ \propto\tilde G_3(0)^n$, is nonvanishing as
 well. Hence, the above contribution to $\tilde F_n$ is relevant in
 the critical region, and should be taken into account. By $\tilde
 F^{0odd}_n({\bf k}_1,...,{\bf k}_n)=F^{0odd}_{n}+ O(k^2)$ we denote
 the contribution to $\tilde F^{odd}_n({\bf k}_1,...,{\bf k}_n)$ which
 is given by all odd secondary diagrams such that no pair of vertices
 is connected only by a chain of the $g$-hyperloops and no loops
 representing ${\cal G}$ are present. The vertex part $\tilde
 F^{0odd}_n({\bf 0},...,{\bf 0})= F^{0odd}_{n}$ is relevant in the
 critical region, and $F^{0odd}_n$ should be taken into account in the
 same way as the vertices ${\cal A}_{2n}$ are. Thus, the contributions
 to any vertex function, which are relevant in the critical region,
 consist of the even secondary diagrams, as well as of the relevant
 odd secondary diagrams (those which contain the dashed lines). Any
 relevant 1PI secondary diagram contains vertices ${\cal A}_6$, ${\cal
 A}_8$ and the vertex parts $F^{0odd}_3$, $F^{0odd}_4$, from which
 there emanate 3 or 4 linear chains of hyperloops respectively, and
 the chains are of the form shown in Fig.17.  An example of an odd secondary
 diagram contributing to the four-point connected correlation function
 is shown in Fig.19.

As in the case of the even diagrams, we can segregate all the relevant
diagrams (both even and odd) into disjoint sets. To a given set there
belongs a diagram in which particular pairs of vertices or the vertex
parts are connected by the linear chains of the hyperloops. All
secondary diagrams obtained from this particular diagram by changing
the number of the hyperloops in the chains belong to the same
set. This is analogous to the segregation of the even diagrams
described in sec.VD.
\begin{figure}
\includegraphics[scale=0.45]{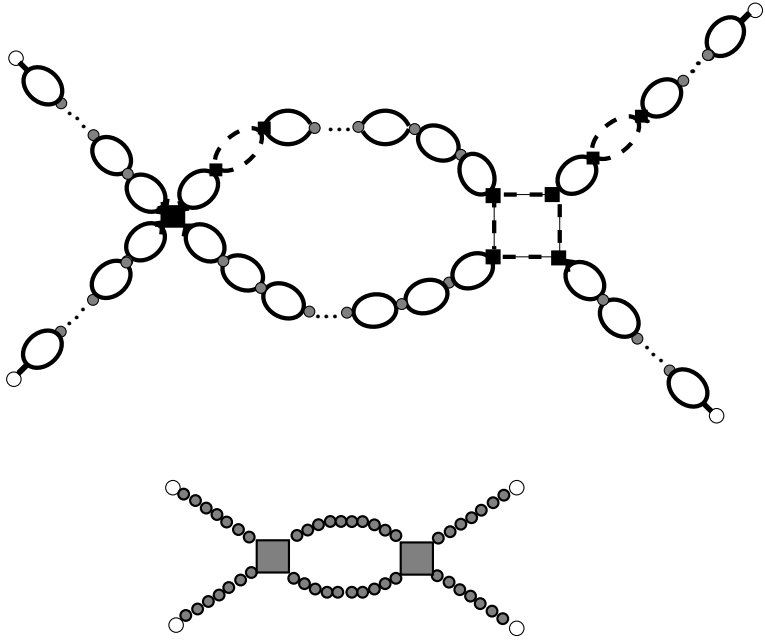}
\caption{Top: A relevant odd secondary diagram contributing to the 
 four-point connected correlation function for the field $\psi$. The
 gray circles, black squares, open circles and black rectangles
 represent ${\cal A}_6$, ${\cal A}_8$, external points, and $a_1$
 respectively. The solid and dashed lines represent the two- and the
 three-point correlation functions for the field $\phi$,
 $G_{\phi\phi}$ and $G_3$ respectively. All diagrams obtained by
 changing the number of the solid- and the dashed-line loops in the
 four linear chains belong to the same set of diagrams. Bottom: A
 hyperdiagram obtained after the resummation over all numbers $n\ge 1$
 of the solid-line loops and over $0\le m\le n-1$ dashed-line loops in
 each linear chain, and over all subdiagrams contributing to the bare
 four-point vertex functions. The pearl-line represents the series of
 all linear chains and the bare four-point vertex functions are
 represented here as the shaded squares. See text for more datails.}
\end{figure}
 There is one-to-one
 correspondence between the vertex functions in our theory and the
 vertex functions obtained in the theory with the coarse-grained
 Hamiltonian (\ref{calH}), when  the terms
\begin{equation}
\frac{1}{3!}\int_{{\bf k}_1}...\int_{{\bf k}_3}(2\pi)^d
\delta(\sum_i^3{\bf k}_i)
\tilde F^{0odd}_{3}({\bf k}_1,...,{\bf k}_3)
\prod_i^3\tilde\psi({\bf k}_i)
\end{equation}
 and 
\begin{equation}
\frac{1}{4!}\int_{{\bf k}_1}...\int_{{\bf k}_4}(2\pi)^d
\delta(\sum_i^4{\bf k}_i)
\tilde F^{0odd}_4({\bf k}_1,...,{\bf k}_4)\prod_i^4\tilde\psi({\bf k}_i)
\end{equation}
 are added to ${\cal
H}_{int}$ (Eq.(\ref{Hin})). Because of the form of $\tilde
F^{0odd}_n({\bf k}_1,...,{\bf k}_n)$, we can separate the relevant
contributions, 
\begin{equation}
\frac{F^{0odd}_{3}}{3!}\int_{{\bf k}_1}...\int_{{\bf k}_3}
(2\pi)^d
\delta(\sum_i^3{\bf k}_i)
\prod_i^3\psi({\bf k}_i)=
\frac{F^{0odd}_{3}}{3!}\int_{{\bf x}}\psi({\bf x})^3
\end{equation}
 and 
\begin{equation}
\frac{F^{0odd}_4}{4!}\int_{{\bf k}_1}...\int_{{\bf k}_4}(2\pi)^d
\delta(\sum_i^4{\bf k}_i)
\prod_i^4\tilde\psi({\bf k}_i)=\frac{F^{0odd}_4}{4!}
\int_{{\bf x}}\psi({\bf x})^4.
\end{equation}
 In the remaining contributions the integrands are proportional to
 $O(k_i^2)$ for $k_i\to 0$. Such contributions to ${\cal H}_{int}$ are
 irrelevant in the RG sense~\cite{zinn-justin:89:0}.  The vertex
 functions for the field $\psi$ scale in the same way as the vertex
 functions in the model given by the Hamiltonian (\ref{calH}) and
 (\ref{Hin}), but with all the parameters modified according to the
 above discussion. In particular, $F^0_n
\rightarrow {\cal A}_n/a_1^n+ F^{0odd}_n$.

  With higher-order vertices ${\cal A}_{2m}$, $m>4$ included,
similar vertex parts with loops made of subdiagrams contributing to
$\langle\phi^5({\bf x}_1)\psi^5({\bf x}_2)\rangle^{con}$ etc. would
contribute to $F^0_3$, $F^0_4$ in Eq. (\ref{Hin}), and to the irrelevant
operators. Modification of the coupling constants affects only the
nonuniversal properties of the theory. Corrections to scaling
associated with irrelevant operators have been studied before
\cite{zinn-justin:89:0}, and are not specific for the RPM.

Let us summarize the above results. From all the secondary diagrams
contributing to the vertex function for the field $\psi$, $\tilde
F_n({\bf k}_1,...,{\bf k}_n)$, we can distinguish a set of secondary
diagrams contributing to the 'bare' $n$-point vertex function. To the
above set there belong the vertex ${\cal A}_{2n}/a_1^n$ and the odd
secondary diagrams such that no pair of vertices ${\cal A}_{2m}$ is
connected by the linear chain of the hyperloops representing $g$
(including the one-loop chains representing $g^H$). Also, no loops
representing ${\cal G}$ are present in the diagrams contributing to
the bare vertex function. The bare vertex function has the form
$\tilde F^{bare}_n({\bf k}_1,...,{\bf k}_n)={\cal
A}_{2n}/a_1^n+F^{0odd}_n+O(k^2)$ for $k\to 0$, where $F^{0odd}_n$ is a
real number.  The remaining secondary diagrams contributing to the
vertex function $\tilde F_m({\bf k}_1,...,{\bf k}_m)$ consist of the
subdiagrams of the form described above, and these subdiagrams are
connected by the chains of the hyperloops representing $g$ (including
the one-loop chains representing $g^H$).  All the secondary diagrams
contributing to the $n$-point vertex function for the field $\psi$ can
be segregated into disjoint sets. Each set contains all the secondary
diagrams obtained from one particular secondary diagram by changing
the number of the hyperloops in the chains connecting the subdiagrams
that belong to the set representing the bare vertex functions. Also
the secondary diagrams obtained from the chosen diagram by changing
the subdiagrams that contribute to the bare vertex functions belong to
the considered set. All the secondary diagrams in any given set are of
the same topological structure when the linear chains are replaced by
lines, and the subdiagrams contributing to the bare $n$-point vertex
function are replaced by vertices from which $n$
lines  emanate. The series of all the secondary diagrams in a given set can be
represented by a hyperdiagram. The hyperdiagram has the same
topological structure as the secondary diagrams in the set. In the
hyperdiagrams the functional form of the line representing the sum
over all numbers of the hyperloops in the linear chain is $1/(c_0+c_2k^2)$ in
Fourier representation, and the $n$-point hypervertices are of the
form ${\cal A}_{2n}/a_1^n+F^{0odd}_n+O(k^2)$ for $k\to 0$. The
generating functional for such vertex functions is obtained from the
Hamiltonian (\ref{calH}) with (\ref{Hin}), that contains also
additional, irrelevant contributions to ${\cal H}_{int}$ that yield
corrections to scaling.
\subsection{Corrections to scaling specific for the RPM}
We found that the connected correlation functions for the field $\psi$
scale in the same way as in the Ising universality class. However, the
correlation functions for the number-density, given in
Eq.(\ref{etetcon1}), contain also other contributions. In particular,
the one- and two-point number-density correlation functions in the RPM
assume the forms
\begin{eqnarray}
\label{roW}
\langle\eta({\bf r})\rangle=
 \langle\psi({\bf r})\rangle 
+
\sum_{n\ge 1}\frac{a_n}{a_1^nn!}\langle\psi^n({\bf r})
\rangle^{con}
\end{eqnarray}
and
\begin{eqnarray}
\label{corWF}
\langle\eta({\bf r}_1)\eta({\bf r}_2)\rangle^{con}=
 \langle\psi({\bf r}_1)\psi ({\bf r}_2)\rangle^{con} 
+\sum_{n,m\ge 1}\sum_{n+m>2}
\frac{a_na_m}{a_1^{n+m}n!m!}\langle\psi^n({\bf r}_1)
\psi^m ({\bf r}_2)\rangle^{con}
\end{eqnarray}
respectively, where the averages are obtained with the Hamiltonian
(\ref{hef}). 
In Figs.20 and 21 we show the secondary
diagrams contributing to the next-to-leading order terms in the above
expansions. On the right the hyperdiagrams
obtained after the resumation of all such diagrams are shown.
\begin{figure}
\includegraphics[scale=0.4]{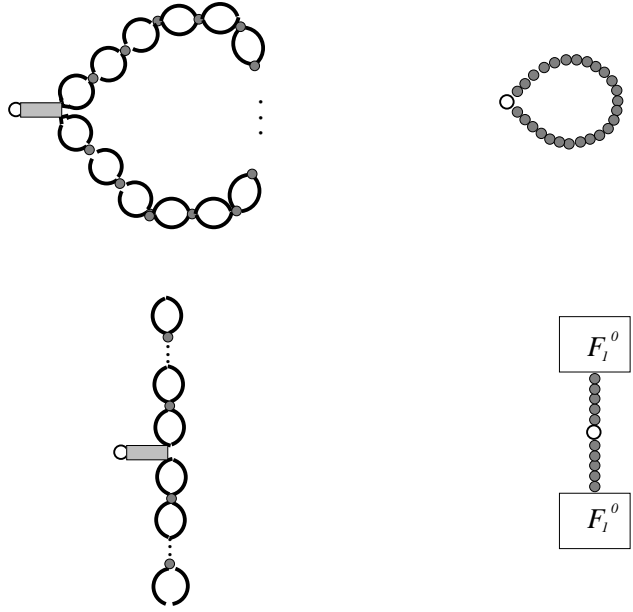}
\caption{Left: Secondary  diagrams contributing to
 the next-to-leading order term 
in the expansion of $\langle\zeta({\bf r})\rangle$ (Eq.(\ref{roW})) at
the zeroth order in the hypervertices ${\cal A}_{2n}$ with $n>2$.  The
external point is shown as an open circle,  gray boxes
represent $a_2/2$, lines represent
$G_{\phi\phi}$ and gray circles represent ${\cal A}_4$. Right:
Hyperdiagrams resulting from the resummation of all diagrams of the
type shown on the left.}
\end{figure}
\begin{figure}
\includegraphics[scale=0.4]{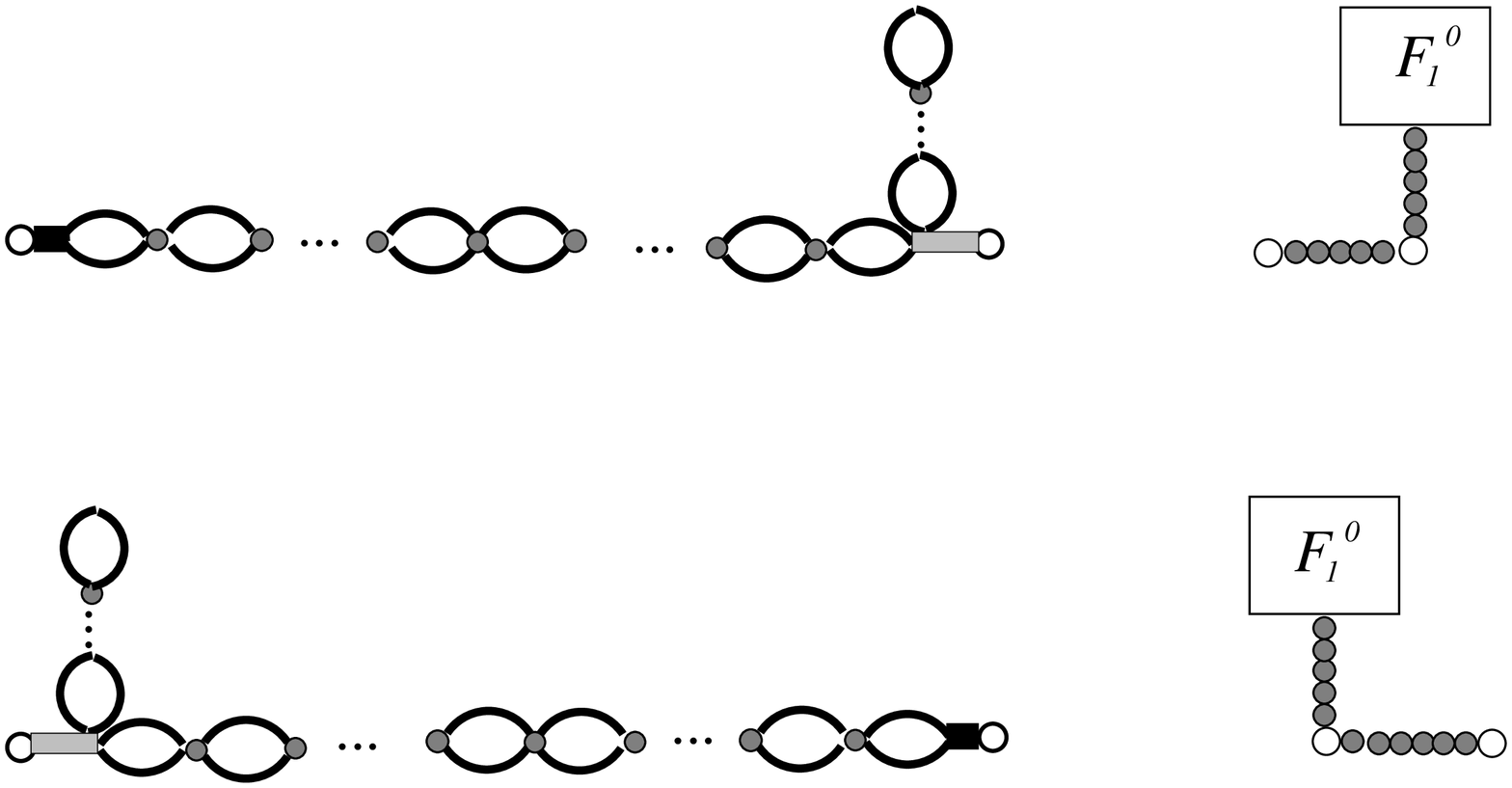}
\caption{Left: secondary  diagrams contributing to the
 next-to-leading order term in the expansion of $G_{\zeta\zeta}$
 (Eq.(\ref{corWF})) at the zeroth order in the hypervertices ${\cal
 A}_n$ with $n>4$.  The external points are shown as open
 circles. black and gray boxes represent $a_1$ and $a_2/2$
 repsectively, lines represent $G_{\phi\phi}$ and gray circles
 represent ${\cal A}_4$. Right: Hyperdiagrams resulting from the
 resummation of all diagrams of the type shown on the left.}
\end{figure}
 In general, the leading-order correction to the $N$-point connected
 correlation function for the field $\psi$ is given by the $N+1$-point
 connected correlation function, where two of the external points 
are identified. 
Scaling forms of the connected correlation
 functions, Eq.  (\ref{Issc}), give
\begin{eqnarray}
G^R_N({\bf r}_1,...,{\bf r}_N;t,u,H)=|t|^{N\beta}
G_{Ns}({\bf r}_1/\xi,...,{\bf r}_N/\xi;H|t|^{-\Delta})+\\
\nonumber
|t|^{(N+1)\beta}
G_{(N+1)s}({\bf r}_1/\xi,{\bf r}_1/\xi,...,{\bf r}_N/\xi;H|t|^{-\Delta})
+O(|t|^{(N+2)\beta}),
\end{eqnarray}
where $G_{Ns}({\bf r}_1/\xi,...,{\bf r}_N/\xi;H|t|^{-\Delta})$ is the
scaling function.   Consider now the correction term to the grand
potential. Because the relative correction to any connected
correlation function is $\propto |t|^{\beta}$, the relative correction
to their generating functional has the same scaling form, and we
obtain
\begin{eqnarray}
\Omega(t,H)=|t|^{2-\alpha}\Big[\omega_s(H|t|^{-\Delta})+
 |t|^{\beta}\omega_{cs}(H|t|^{-\Delta})+...\Big].
\end{eqnarray} 
All derivatives of the grand potential can be written in the form
\begin{eqnarray}
X\propto |t|^a\Big[1+ |t|^{\beta}X_{cs}(H|t|^{-\Delta})+O(|t|^{2\beta})\Big],
\end{eqnarray} 
where $a$ is the corresponding critical parameter in the Ising
universality class. The leading-order correction to scaling, specific
for the RPM, is given by the exponent $\beta$.
\section{summary}
The theory outlined above is relatively complex compared to the theory
of critical phenomena in simple systems. This complexity follows from
the indirect nature of criticality in ionic systems. The long-range
critical number-density fluctuations in charged systems are not
induced by mechanical interactions. Unlike in systems with short-range
interactions, the phase separation is induced by strong charge-density
correlations. Coulombic forces support formation of charge-ordered
structures with a microscopic distance between oppositely charged
neighbors. Individual microscopic states are charge-ordered with a
high probability as predicted theoretically \cite{ciach:01:0} and seen
in snapshots \cite{weis:98:0,panag:02:0,yan:02:0}, but fluctuations
restore the uniform structure at larger length scales (and large
observation times). Mathematically the restored disorder is described
within the Brazovskii approach. In our case we performed a resummation
of singular Feynman diagrams (singularity resulting from
charge-ordering) to obtain regular secondary diagrams. The
charge-ordered 'living' clusters that are formed in different
microstates interact with each other with short-range forces, and this
observation suggests standard criticality. In microscopic description
interactions between clusters of various sizes, shapes and
orientations should be considered.  The above complexity in our theory
is reflected in the absence of critical singularity in individual
secondary diagrams. The critical instability of the uniform phase
occurs only when an infinite series of secondary diagrams contributing
to the correlation function $\langle\phi^2({\bf x}_1)\phi^2({\bf
x}_2)\rangle^{con}$ is calculated in the perturbation expansion. In
the diagrams of a form of chains of $n$ loops the correlations with
$n-1$ intermediate points are included. The series of chains of loops
plays an analogous role as the Gaussian correlation function in simple
fluids. The hyperdiagrams with the lines representing such series play
in turn an analogous role as diagrams in the standard theory of
critical phenomena.

From the point of view of the field theory, the above work concerns
the theory with the action of the form (\ref{hef}) with ${\cal A}_4<0$
and ${\cal A}_{2n}>0$ for $n>3$, and in Fourier space the action
becomes unstable for the wavenumber $k_b\ne 0$. The key properties
${\cal A}_4<0$ and ${\cal A}_{2n}>0$ have been observed within the WF
approximation at low densities for the RPM with the ideal entropy of
mixing \cite{ciach:04:1} and with the
Percus-Yevick \cite{patsahan:05:0} reference system. It is plausible
that for the exact form of ${\cal A}_4$ the condition ${\cal A}_4<0$
is satisfied in the RPM for low densities, but it remains to be
proven. This work shows that the model for which the above is
satisfied belongs to the Ising universality class.

\begin{acknowledgments}
I greatly benefited from  inspiring  
discussions with Professor Stell, especially at the early stage of this work. 
 This work was  supported
by the KBN  through the research project No. 1 P02B 033 26.
\end{acknowledgments}


\begin{thebibliography}{10}

\bibitem{stell:76:0}
G. Stell, K. Wu, and B. Larsen, {\it Phys. Rev. Lett.} {\bf 37},  1369  (1976).

\bibitem{hafskjold:82:0}
B. Hafskjold and G.Stell,  in {\em The Liquid State of matter. Fluids, Simple
  and Complex}, edited by E. Montroll and J. Lebowitz (North Holland,
  Amsterdam, 1982).

\bibitem{stell:92:0}
G. Stell, {\it Phys. Rev. A} {\bf 45},  7628  (1992).

\bibitem{stell:95:0}
G. Stell, {\it J. Stat. Phys.} {\bf 78},  197  (1995).

\bibitem{ciach:00:0}
A. Ciach and G. Stell, {\it J. Mol. Liq.} {\bf 87},  253  (2000).

\bibitem{ciach:02:0}
A. Ciach and G. Stell, {\it Physica A} {\bf 306},  220  (2002).

\bibitem{ciach:05:0}
A. Ciach and G. Stell, {\it Int.J. Mod. Phys. B} {\bf 21},  3309  (2005).

\bibitem{japas:90:0}
M.~L. Japas and J.~M.~H. Levelt-Sengers, {\it J. Phys. Chem.} {\bf 94},  5361
  (1990).

\bibitem{narayanan:94:0}
T. Narayanan and K.~S. Pitzer, {\it J. Phys. Chem.} {\bf 98},  9170  (1994).

\bibitem{pitzer:90:0}
K.~S. Pitzer, Acc. Chem. Res. {\bf 23},  373  (1990).

\bibitem{singh:90:0}
R.~R. Singh and K.~S. Pitzer, {\it J. Chem. Phys.} {\bf 92},  6775  (1990).

\bibitem{zhang:92:0}
K.~C. Zhang, M.~E. Briggs, R.~W. Gammon, and J.~M.~H. Levelt-Sengers, {\it J.
  Chem. Phys.} {\bf 97},  8692  (1992).

\bibitem{weingaertner:01:0}
H. Weing\"artner and W. Schr\"oer, Adv. Chem. Phys {\bf 116},  1  (2001).

\bibitem{bianchi:01:0}
H.~L. Bianchi and M.~L. Japas, {\it J. Chem. Phys.} {\bf 115},  10472  (2001).

\bibitem{anisimov:00:0}
M. Anisimov {\it et~al.}, {\it Phys. Rev. Lett.} {\bf 85},  2336  (2000).

\bibitem{dickman:99:0}
R. Dickman and G. Stell,  in {\em Simulation and theory of electrostatic
  interactions in solution}, edited by L. Pratt and G. Hummer (AIP Conf.
  Proceedings 492, Melville, NY, 1999).

\bibitem{panag:99:0}
A.~Z. Panagiotopoulos and S. Kumar, {\it Phys. Rev. Lett.} {\bf 83},  2981
  (1999).

\bibitem{diehl:03:0}
A.Diehl and A.Z.Panagiotopoulos, {\it J. Chem. Phys.} {\bf 118},  4993  (2003).

\bibitem{hoye:97:0}
J. Hoye and G. Stell, {\it J. Stat. Phys.} {\bf 89},  177  (1997).

\bibitem{stell:99:0}
G. Stell,  in {\em New Approaches to Problems in Liquid-State Theory}, edited
  by C. Caccamo, J.-P. Hansen, and G. Stell (Kluwer Academic Publishers,
  Dordrecht, 1999).

\bibitem{ciach:01:1}
A. Ciach and G. Stell, {\it J. Chem. Phys.} {\bf 114},  3617  (2001).

\bibitem{kobelev:02:0}
V. Kobelev, A.~B. Kolomeisky, and M. Fisher, {\it J. Chem. Phys.} {\bf 117},
  8897  (2002).

\bibitem{brognara:02:0}
A.Brognara, A. Parola, and L. Reatto, {\it Phys. Rev. E} {\bf 65},  66113
  (2002).

\bibitem{fisher:94:0}
M.~E. Fisher, {\it J. Stat. Phys.} {\bf 75},  1  (1994).

\bibitem{wiegand:94:0}
S. Wiegand {\it et~al.}, Int. J. Thermophys. {\bf 15},  1045  (1994).

\bibitem{wiegand:97:0}
S. Wiegand {\it et~al.}, {\it J. Chem. Phys.} {\bf 106},  2777  (1997).

\bibitem{wiegand:98:0}
S. Wiegand {\it et~al.}, {\it J. Chem. Phys.} {\bf 109},  9038  (1998).

\bibitem{wiegand:98:1}
S. Wiegand, R.~F. Berg, and J.~M. Levelt-Sengers, {\it J. Chem. Phys.} {\bf
  109},  4533  (1998).

\bibitem{kleemeier:99:0}
M. Kleemeier, S. Wiegand, W. Schr\"oer, and H. Weing\"artner, {\it J. Chem.
  Phys.} {\bf 110},  3085  (1999).

\bibitem{wagner:02:0}
M. Wagner, O. Stanga, and W. Schr\"oer, {\it PCCP} {\bf 4},  5300  (2002).

\bibitem{wagner:03:0}
M. Wagner, O. Stanga, and W. Schr\"oer, {\it PCCP} {\bf 5},  3943  (2003).

\bibitem{wagner:04:0}
M. Wagner, O. Stanga, and W. Schr\"oer, {\it PCCP} {\bf 6},  580  (2004).

\bibitem{kostko:04:0}
A.~F. Kostko, M.~A. Anisimov, and J.~V. Sengers, {\it Phys. Rev. E} {\bf 70},
  026118  (2004).

\bibitem{orkoulas:99:0}
G. Orkoulas and A.~Z. Panagiotopoulos, {\it J. Chem. Phys.} {\bf 110},  1581
  (1999).

\bibitem{yan:99:0}
Q. Yan and J.~J. de~Pablo, {\it J. Chem. Phys.} {\bf 111},  9509  (1999).

\bibitem{luijten:01:0}
E. Luijten, M. Fisher, and A. Panagiotopoulos, {\it J. Chem. Phys.} {\bf 114},
  5468  (2001).

\bibitem{caillol:02:0}
J.-M. Caillol, D. Levesque, and J.-J. Weis, {\it J. Chem. Phys.} {\bf 116},
  10794  (2002).

\bibitem{luijten:02:0}
E. Luijten, M. Fisher, and A. Panagiotopoulos, {\it Phys. Rev. Lett.} {\bf 88},
   185701  (2002).

\bibitem{kim:03:0}
Y.~C. Kim, M.~E. Fisher, and E. Luijten, {\it Phys. Rev. Lett.} {\bf 91},
  065701  (2003).

\bibitem{amit:84:0}
D.~J. Amit, {\em Field Theory, the Renormalization Group and Critical
  Phenomena} (World Scientific, Singapore, 1984).

\bibitem{zinn-justin:89:0}
Zinn-Justin, {\em Quantum Field Theory and Critical Phenomena} (Clarendon
  Press, Oxford, 1989).

\bibitem{ciach:05:2}
A. Ciach, W.~T. Gozdz, and G. Stell, cond-mat/0411424  .

\bibitem{brazovskii:75:0}
S.~A. Brazovskii, {\it Sov. Phys. JETP} {\bf 41},  8  (1975).

\bibitem{ciach:04:1}
A. Ciach, {\it Phys. Rev. E} {\bf 70},  046103  (2004).

\bibitem{ciach:03:0}
A. Ciach and G. Stell, {\it Phys. Rev. Lett.} {\bf 91},  60601  (2003).

\bibitem{ciach:04:0}
A. Ciach and G. Stell, {\it Phys. Rev. E} {\bf 70},  16114  (2004).

\bibitem{diehl:05:0}
A. Diehl and A.~Z. Panagiotopolous, {\it Phys. Rev. E} {\bf 71},  046118
  (2005).

\bibitem{fredrickson:87:0}
G.~H. Fredrickson and E. Helfand, {\it J. Chem. Phys.} {\bf 87},  67  (1987).

\bibitem{ciach:01:0}
A. Ciach and G. Stell, {\it J. Chem. Phys.} {\bf 114},  382  (2001).

\bibitem{weis:98:0}
J.-J. Weis, D. Levesque, and J. Caillol, {\it J. Chem. Phys.} {\bf 109},  7486
  (1998).

\bibitem{panag:02:0}
A. Panagiotopoulos and M. Fisher, {\it Phys. Rev. Lett.} {\bf 88},  45701
  (2002).

\bibitem{yan:02:0}
Q. Yan and J. de~Pablo, {\it Phys. Rev. Lett.} {\bf 88},  95504  (2002).

\bibitem{patsahan:05:0}
O. Patsahan and A. Ciach,   , to be published.

\end{thebibliography}

\end{document}